\documentclass[aps,floats,amssymb,amsmath,prl,nofootinbib,twocolumn,superscriptaddress]{revtex4-1}

\usepackage{booktabs}
\usepackage{calc}
\usepackage{psfrag}
\usepackage{graphicx}
\usepackage{color}
\usepackage{units}
\usepackage[dvipsnames]{xcolor}
\usepackage[unicode=true,pdfusetitle,
 bookmarks=true,bookmarksnumbered=false,bookmarksopen=false,
 breaklinks=false,pdfborder={0 0 0},backref=false,colorlinks=true]
 {hyperref}
\usepackage{geometry}
\pdfpageattr {/Group << /S /Transparency /I true /CS /DeviceRGB>>} 

\newcommand{\up}{\uparrow}
\newcommand{\dn}{\downarrow}

\usepackage{pdfpages}

\makeatletter
\AtBeginDocument{\let\LS@rot\@undefined}
\makeatother

\begin{document}

\title{Towards numerically exact computation of conductivity in the thermodynamic limit of interacting lattice models}

\author{Jeremija Kovačević}

\affiliation{Scientific Computing Laboratory, Center for the Study of Complex Systems,\\
Institute of Physics Belgrade, University of Belgrade, Pregrevica 118,
11080 Belgrade, Serbia}

\author{Michel Ferrero}

\affiliation{CPHT, CNRS, Ecole Polytechnique, Institut Polytechnique de Paris, Route de Saclay, 91128 Palaiseau, France}
\affiliation{Coll\`ege de France, 11 place Marcelin Berthelot, 75005 Paris, France}

\author{Jakša Vučičević}

\affiliation{Scientific Computing Laboratory, Center for the Study of
Complex Systems,\\
Institute of Physics Belgrade, University of Belgrade, Pregrevica 118,
11080 Belgrade, Serbia}

\begin{abstract}
Computing dynamical response functions in interacting lattice models is a long standing challenge in condensed matter physics.
In view of recent results, the dc resistivity $\rho_\mathrm{dc}$ in the weak coupling regime of the Hubbard model is of great interest, yet it is not fully understood.
The challenge lies in having to work with large lattices while avoiding analytical continuation.
The weak-coupling $\rho_\mathrm{dc}$ results were so far computed at the level of the Boltzmann theory and at the level of the Kubo bubble approximation, which neglects vertex corrections. Neither theory was so far rigorously proven to give exact results even at infinitesimal coupling, and the respective dc resistivity results differ greatly.
In this work we develop, cross-check and apply two state-of-the-art methods for obtaining dynamical response functions. We compute the optical conductivity at weak coupling in the Hubbard model in a fully controlled way, in the thermodynamic limit and without analytical continuation. We show that vertex corrections persist to infinitesimal coupling, with a constant ratio to the Kubo bubble. We connect our methods with the Boltzmann theory, and show that the latter applies additional approximations, which lead to quantitatively incorrect scaling of $\rho_\mathrm{dc}$ with respect to the coupling constant.
\end{abstract}

\maketitle

Strongly correlated electronic systems often display rich, yet remarkably universal phase diagrams\cite{Keimer2015,Limelette2003,Kurosaki2005,Kagawa2005,Dumm2009,Powell2011,Furukawa2015,Andrei2020,Cao2020,Li2021}.
One of the most puzzling universal phenomena is the strange-metallic linear-in-temperature dc resistivity\cite{Takagi1992,Cooper2009,Legros2018,KrishnaKumar2017,KrishnaKumar2018,Cao2020,Wang2020,Grissonnanche2021,Jaoui2022,Ayres2021}.
It appears in unconventional and high-temperature superconductors, in the regime where their critical temperature $T_c$ is the highest\cite{Keimer2015,Takagi1992,Cooper2009,Wang2020,Ayres2021}.
In other cases, strange metals are associated with quantum critical points\cite{Grigera2001,Cao2020,Grissonnanche2021,VucicevicPRL2015,ChaPNAS2020}. This raises the question whether there is an intimate connection between criticality,
transport properties and the magnitude of the superconducting $T_c$. To make sense of the vast experimental data, one must be able to compute the conductivity in interacting lattice models, which is a difficult, long standing task.
The main challenge is to find a way to obtain controlled results on the real frequency axis and, at the same time, avoid finite lattice-size effects.
Exact diagonalization based methods [finite-temperature Lanczos (FTLM)\cite{KokaljPRBR2017,VucicevicPRL2019,Vasic2024}, linked cluster expansions (NLCE)\cite{Rigol2006,Khatami2011,Tang2013}], and the density-matrix renormalization group (DMRG)\cite{Qu2024} are all inherently limited to small lattice sizes. Quantum Monte Carlo methods, on the other hand, either require analytical continuation\cite{Denteneer1999,HuangScience2019,SimkovicScience2024} or are effectively limited to atomic problems\cite{Schiro2010,Cohen2015,Profumo2015,Machek2020,Bertrand2021}. In the special case of Hall resistivity, expansions in terms of thermodynamic quantities allow for progress\cite{Auerbach2019,Khait2023}. In this paper, however, we formulate a general and systematic way forward.

The workhorse model for the description of the cuprates (and many other classes of correlated systems) is the Hubbard model\cite{Qin2022,Limelette2003,VucicevicPRL2011,Furukawa2015,Li2021,Deng2014,VucicevicPRL2015,VucicevicPRL2021,KrishnaKumar2017}. Early works\cite{VucicevicPRL2015} have shown that the inifinite-dimensional Bethe-lattice Hubbard model roughly describes the normal phase resistivity in LSCO at moderate to high temperature. However, the physics at low temperature is expected to be dominated by the dimensionality of the model, and thus of primary interest is the Hubbard model on the 2D square-lattice. At very strong coupling and high temperature, small 2D lattices become representative of the thermodynamic limit, and FTLM was used to obtain numerically exact results\cite{VucicevicPRL2019,KokaljPRBR2017}. However, to address the questions of strange metallic behavior and its connection to quantum critical points\cite{Keimer2015, ChaPNAS2020, Jaoui2022, Cooper2009,Legros2018,Wang2020,Grigera2001}, one must be able to perform computations at lower temperature and, perhaps, lower coupling, a regime where small-cluster methods fail.

Recent works\cite{XuPRR2022, SimkovicScience2024} have indicated that the ground-state phase diagram of the (nearest-neighbor hopping) square-lattice Hubbard model features a quantum critical line (QCL), delineating an ordered stripe ground state. The QCL passes through zero coupling at zero doping (i.e.  half-filling). At this point, charge and spin susceptibility diverge\cite{VucicevicPRB2023}, and both the Boltzmann theory\cite{KielyPRB2023,Schneider2012} and the Kubo bubble\cite{VucicevicPRB2023} predict a linear-in-temperature resistivity down to the lowest accessible temperature. This finding is in line with numerous observations of linear resistivity in the vicinity of quantum critical points\cite{ChaPNAS2020, Jaoui2022,Cooper2009,Legros2018,Wang2020,Grigera2001}. \emph{Kiely and Muller}\cite{KielyPRB2023} have argued that the linear-resistivity strange metal observed at half-filling/weak coupling is connected to the strange metal in the cuprates, corresponding to the strong coupling/finite doping regime of the Hubbard model.

However, our recent results\cite{VucicevicPRB2023} have shown a strong quantitative disagreement between Boltzmann theory and the Kubo bubble, casting doubt on whether either of the theories captures correctly even the qualitative behavior of resistivity. To resolve the phenomenology at weak coupling, better methods are needed.


In this work, we address the conductivity in the square lattice Hubbard model.
We develop two state-of-the-art methodologies and fully avoid finite-size effects and the uncontrolled analytical continuation\cite{VucicevicPRL2019,HuangScience2019,BergeronPRB2011}.

First, we make use of the real-frequency diagrammatic Monte Carlo (RFDiagMC)\cite{VucicevicPRR2021,VucicevicPRB2020,TaheridehkordiPRB2019,GrandadamArxiv2023}, which relies on constructing a power-series expansion for a given physical quantity; the resulting Feynman diagrams are computed up to a given order and then the series is (re)summed. The imaginary-time integrals in Feynman diagrams are solved analytically (which circumvents analytical continuation), while spatial degrees of freedom are summed over using (quasi) Monte Carlo\cite{Sobol1967, Machek2020, Strand2024}. The thermodynamic limit is treated directly.

Next, we devise three different non-equilibrium protocols, where we perturb the system with small external fields and compute the current response as a function of time; we then use the results to reconstruct the optical and dc conductivity in a manner of ``inverse linear response theory''.
In practice, we solve the Kadanoff-Baym equations to obtain the Green's function, given an approximation for the self-energy as input.
We do this calculation for lattices as large as $60\times 60$ and confirm convergence of the results with lattice size.

Our diagrammatic series expansion and the corresponding non-equilibrium results are in excellent agreement, which confirms the validity of both implementations.
As the coupling constant approaches zero, we observe that vertex corrections to dc conductivity do not vanish, but rather diverge with the same power-law scaling as the Kubo bubble contribution, meaning that they remain quantitatively important even at infinitesimal coupling. Vertex corrections are, however, not very big relative to the Kubo bubble. Nevertheless, neither the Kubo bubble approximation nor the Boltzmann equation yield quantitatively correct results, even at infinitesimal coupling.

\emph{Model.} We are treating the square lattice Hubbard model. The Hamiltonian reads
\begin{equation}
 H = -t \sum_{\langle ij \rangle ,\sigma} c^\dagger_{\sigma,i} c_{\sigma, j} -\mu \sum_{\sigma,i}   n_{\sigma,i}+ U\sum_i n_{\up,i} n_{\dn,i}
\end{equation}
where $i,j$ enumerate lattice sites, $c^\dagger/c$ are creation/annihilation operators, $\sigma=\uparrow,\downarrow$ denotes spin, $t$ is the nearest-neighbor hopping-amplitude, set to $t=0.25$. The particle-number operator is denoted $n_{\sigma,i}=c^\dagger_{\sigma,i} c_{\sigma i}$, and $\mu$ is the chemical potential, which is used to tune the average occupancy of the sites. The coupling constant is denoted $U$. In practice, we absorb the Hartree shift in the chemical potential, $\tilde{\mu}=\mu-U\langle n_{i,\sigma}\rangle$, thus $\tilde{\mu}=0$ corresponds to half-filling. We assume $\hbar=e=1$.

\begin{figure*}[t]
\includegraphics[width=0.95\textwidth, trim={0 1cm 0 0.5cm}]{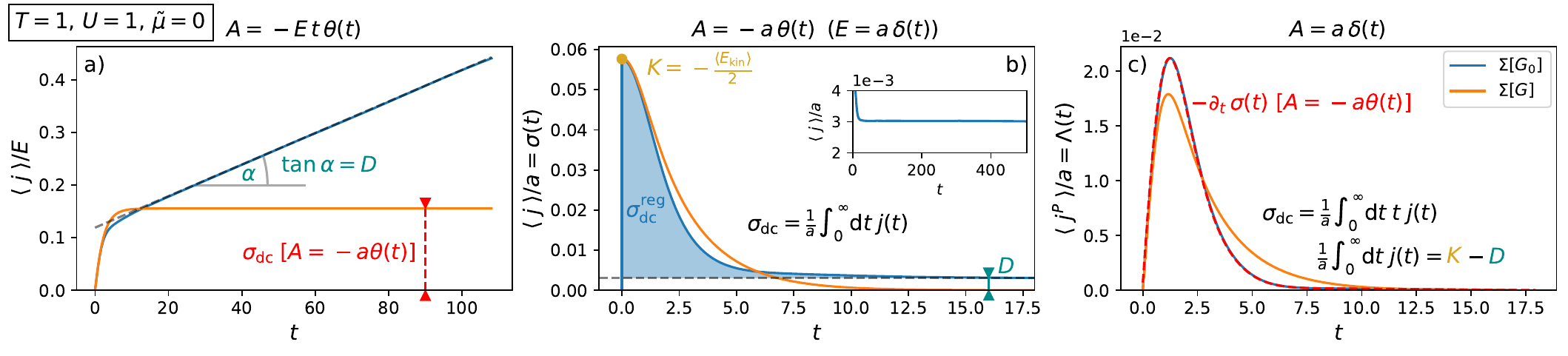}
\centering
\caption{Example of non-equilibrium, inverse linear response theory. Plots show current response vs. time, in three different non-equilibrium protocols: \emph{a}) constant electric field $E$, \emph{b}) short pulse of electric field and \emph{c}) short pulse of vector potential $A$. Protocol \emph{a}) allows to extract $\sigma_\mathrm{dc}$, \emph{b}) and \emph{c}) yield the full $\sigma(t)$ (and thus $\sigma(\omega)$). Different curves correspond to different self-energy approximations, namely $\Sigma[G]$ and $\Sigma[G_0]$. The red dashed lines in panels \emph{a}) and \emph{c}) are comparisons with the protocol \emph{b}). In protocol \emph{c}), we show the paramagnetic part of the current $j^p$ as only this part is relevant. All three protocols yield consistent results. In the $\Sigma[G_0]$ approx. we observe a finite charge-stifness $D$. The inset in panel \emph{b}) zooms in on the long-time tail, showing clearly that $\sigma(t\rightarrow \infty)=D$. }
\label{fig:protocols}
\end{figure*}

\emph{Non-equilibrium approach.} We consider the time-evolution of the Hubbard model which was in a thermal state at times $t<0$, and was then subjected to an external perturbation starting from time $t=0$. Given an approximation for the self-energy, the Green's function can be computed by solving the Kadanoff-Baym equations (we use the code package NESSi\cite{SchuelerCPC2020} and cross check with our own implementation, see supplemental material, SM, for details). Kadanoff-Baym equations are formulated on the three-piece time-contour as\cite{AokiRMP2014}
\begin{equation}
 G(t,t')[-i\overleftarrow{\partial}_{t'}-h(t')] - \int_{\cal C} \mathrm{d}\bar{t} G(t,\bar{t})\Sigma(\bar{t},t') = \delta_{\cal C}
\end{equation}
Here $G$ is the full Green's function, $\Sigma$ is the self-energy, and $h$ is the single-particle Hamiltonian, which introduces an external electric field through the vector potential $\mathbf{A}$, namely $\mathbf{E} = -\partial_t \mathbf{A}$. We restrict ourselves to fields along the $x$-direction (assuming site-positions to be $\mathbf{r}_i=(x_i,y_i)$, with $x_i,y_i\in\mathbb{Z}$) and the corresponding longitudinal response\cite{VucicevicPRB2021}.
The time-diagonal elements in the lesser component of the Green's function contain information about the uniform current, i.e. $\langle j(t) \rangle = -\frac{i}{N}\sum_{\sigma,\mathbf{k}} v_{\mathbf{k}-\mathbf{A}(t)} G^{<}_{\sigma,\mathbf{k}}(t,t)$\cite{AokiRMP2014,AmaricciPRB2012} and $v_\mathbf{k}$ is the $x$-component of the velocity of an electron in the plane-wave state $\mathbf{k}$.

On the other hand, the time-evolution of the current following application of a weak electric field can be computed based on the knowledge of the retarded current-current correlation function in equilibrium\cite{KennesPRB2017}, $\Lambda$, as
\begin{equation}\label{eq:lrt1}
 \langle j(t) \rangle =  \int_{-\infty}^t \Lambda(t-t') A(t') - K A(t)
\end{equation}
with $K=-\frac{\langle E_\mathrm{kin} \rangle}{2}$, i.e. minus the average kinetic-energy per site per spatial dimension. The first term is the paramagnetic part of the current, the second term is the diamagnetic part (see SM for details).
Alternatively, if one knows the optical conductivity $\sigma$, the current-response is computed as
\begin{equation}\label{eq:lrt2}
 \langle j(t) \rangle = \int_{-\infty}^{t} \mathrm{d}t' \sigma(t-t') E(t')
\end{equation}
The current-current correlation function is related to the optical conductivity through $\sigma(t) = K\theta(t) - \int_0^{t} \mathrm{d}t' \Lambda(t') $, or $\partial_t \sigma = -\Lambda$ (for $t > 0$). The optical conductivities in time and frequency domains are connected via Fourier transformation $\sigma(\omega) = \int_{-\infty}^{\infty} \mathrm{d}t e^{i\omega t} \sigma(t)$, and the dc conductivity is simply $\sigma_\mathrm{dc} \equiv \sigma(\omega=0)$.

We devise non-equilibrium protocols that will allow us to invert the linear response Eqs.~\ref{eq:lrt1} and \ref{eq:lrt2} for $\Lambda(t)$ and $\sigma(t)$, compute them based on the current response, and reconstruct $\sigma(\omega)$.
The three protocols are \emph{a}) constant electric field, \emph{b}) short pulse of electric field and \emph{c}) short pulse of vector potential.
The corresponding expressions for the vector potential $A(t)$ are given in Fig.\ref{fig:protocols}.
We use weak fields and make sure we probe the linear response regime (see SM for details).

\emph{Self-energy approximation.} We compute the self-energy perturbatively in powers of $U$, and truncate at second order. The first order self-energy in the Hubbard model is instantaneous (the Hartree shift) and can be absorbed in the single-particle Hamiltonian $h$. What remains to be computed is a single Feynman diagram:
\begin{equation}
  \Sigma_{ij}(t,t')[G] = U^2 G_{ij}(t,t')G_{ij}(t,t')G_{ji}(t',t)
\end{equation}
However, one may still choose to compute the diagram self-consistently or not, i.e. the propagator appearing in the self-energy diagram can be considered to be the fully dressed propagator ($G$), or the bare propagator ($G_0$). The self-consistent approximation corresponds to an approximation of the Luttinger-Ward functional and is guaranteed to respect charge and energy conservation laws. The two approximations for the self-energy must become indistinguishable as $U\rightarrow 0$, but at any finite $U$, they may yield different results.

\emph{Results.} Our non-equilibrium theory is
illustrated on an example in Fig.\ref{fig:protocols}. We find that the three protocols yield perfectly consistent results (e.g. on Figs.\ref{fig:protocols}a and \ref{fig:protocols}c we show in red color the comparison to the protocol \emph{b}) result).
However, the two self-energy approximations lead to drastically different results. Most importantly, the $\Sigma[G_0]$ approximation yields infinite conductivity. This manifests differently in the three different protocols. In the case of constant electric field, this means there is no stationary state and the current keeps growing with time. In the short electric field pulse case, the current does not decay to zero, but to a finite constant instead (as shown on Fig.\ref{fig:protocols}a, the constant is in perfect agreement with the slope of the linear growth of the current in the protocol \emph{a}). This indicates that the infinite conductivity is due to a finite charge sfiffness $D$, which is when the optical conductivity can be separated in two parts as $\sigma(t) = \sigma^\mathrm{reg}(t) + D\theta(t)$, with the regular part $\sigma^\mathrm{reg}(t)$ decaying to zero at long times\cite{KennesPRB2017,KanekoPRB2020}. In frequency domain this means $\mathrm{Re}\sigma(\omega) = \pi D\delta(\omega) + \mathrm{Re}\sigma^\mathrm{reg}(\omega)$. In the short vector potential pulse case, the current does decay to zero, but the charge stiffness can be deduced from the obtained current-current correlation function based on the relation $\int_0^{\infty} \mathrm{d}t \Lambda(t) = K-D$. Regardless of the $\Sigma$ approximation, the optical sum rule $\sigma(t=0^+)=K=\frac{1}{\pi}\int\mathrm{d}\omega\mathrm{Re}\sigma(\omega)$ is satisfied (Fig.\ref{fig:protocols}b, SM). To confirm that our results indicate charge stiffness, rather than a large conductivity, we have studied how $\sigma(\omega)$ changes in the presence of a small fermionic bath (see SM).


\begin{figure}[t]
\includegraphics[width=0.45\textwidth]{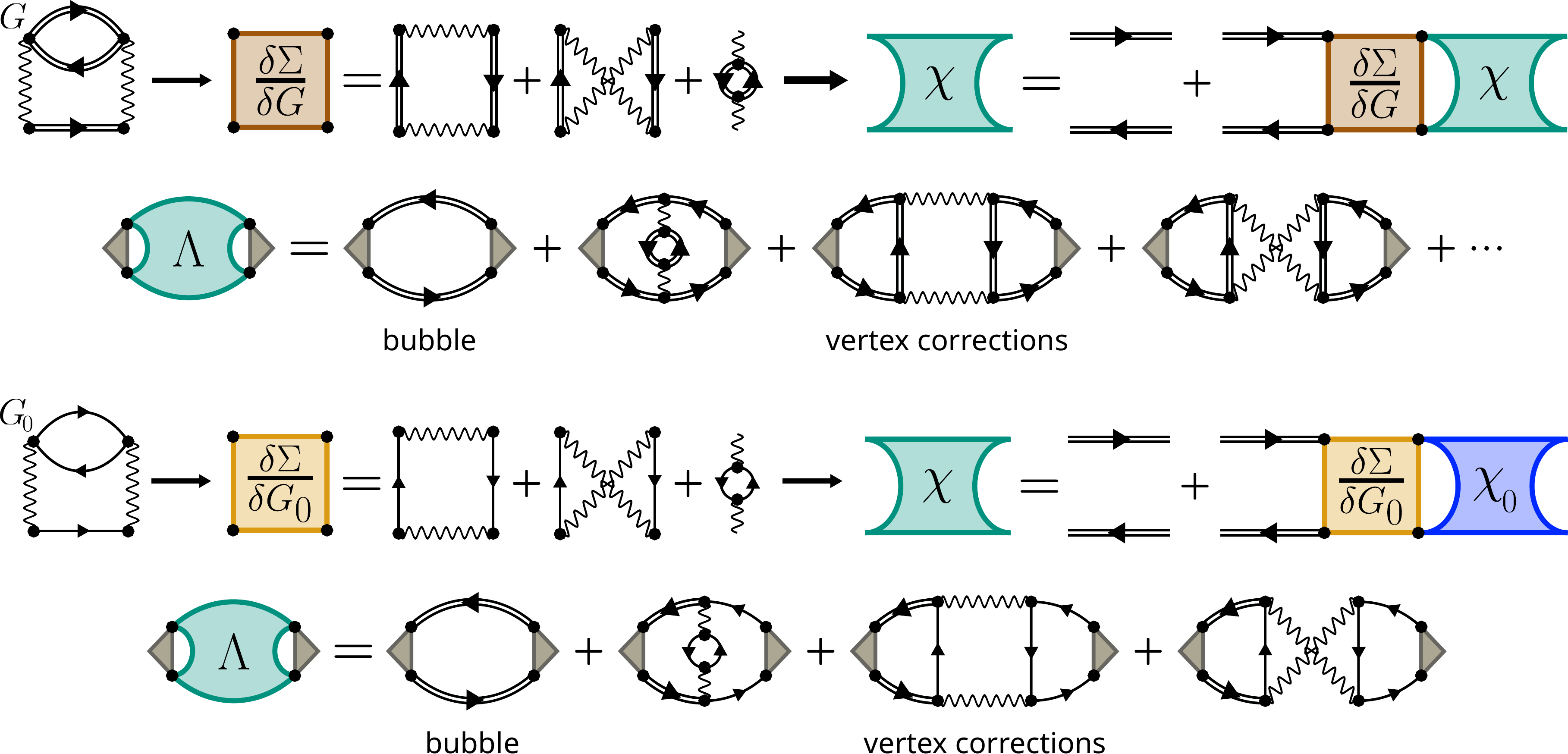}
\centering
\caption{Diagrammatic content of the current-current correlation function effectively computed in our non-equilibrium theory based on different diagrammatic approximations for the self-energy.}
\label{fig:diagrams}
\end{figure}

\begin{figure*}[ht!]
\includegraphics[width=\textwidth,trim={0 1cm 0 1cm}]{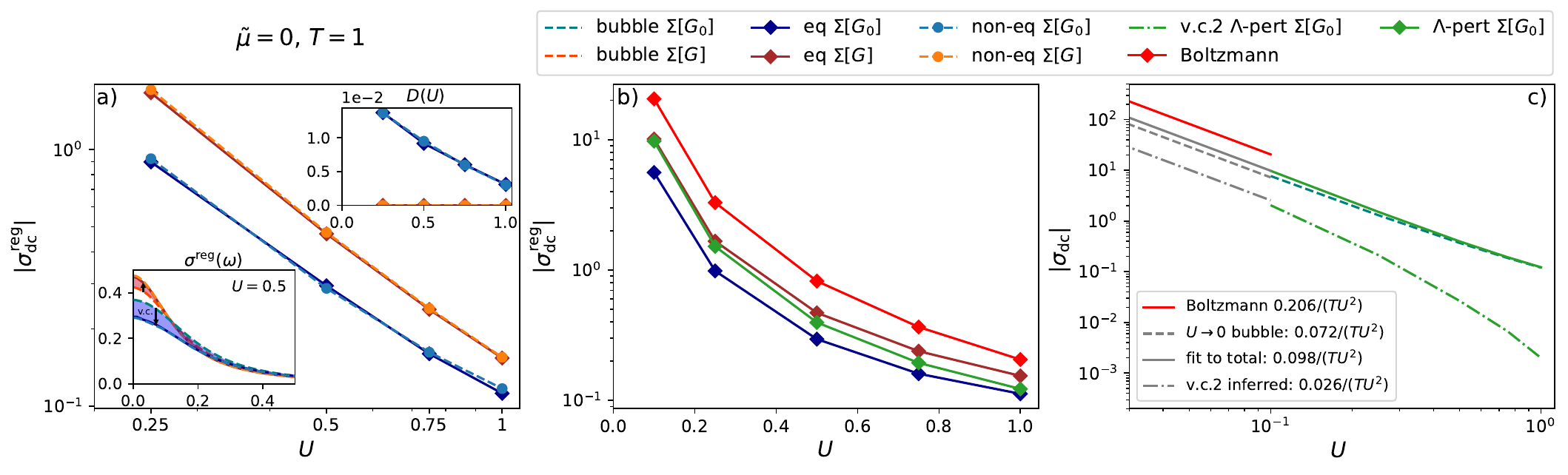}
\centering
\caption{
Main results showing the comparison between different theories and the divergence of vertex corrections in the $U\rightarrow 0$ limit: a) cross-check between equilibrium and the corresponding non-equilibrium theories showing perfect agreement in terms of the regular part of the dc conductivity $\sigma^\mathrm{reg}_\mathrm{dc}$ (main panel), optical conductivity $\mathrm{Re}\sigma^\mathrm{reg}(\omega)$ (lower inset) and the charge-stifness $D$ (upper inset). The lower inset also shows the contribution of the vertex corrections to $\mathrm{Re}\sigma^\mathrm{reg}(\omega)$ (positive in $\Sigma[G]$ approx., negative in $\Sigma[G_0]$ approx., vanishing at high freq.); b) comparison between two possible $\Lambda$-pert. theories (brown and green), the theory consistent with non-eq. $\Sigma[G_0]$ approximation (blue), and the Boltzmann theory(red), showing that different $\Lambda$-pert. theories become indistinguishable as $U\rightarrow 0$, while the non-eq. $\Sigma[G_0]$ and the Boltzman remain different; c) small-$U$ scaling of the results. Gray lines display the strict $U\rightarrow 0$ scaling: the dashed gray line denotes the bubble computed in this limit, using the approach explained in \cite{VucicevicPRB2023}; the full gray line is a fit to the total result (full green line); the dash-dotted is inferred from the previous two; the scaling of Boltzman results is taken from \cite{KielyPRB2023}.}
\label{fig:vertexcorrections}
\end{figure*}

\emph{Crosschecking with RFDiagMC.} To cross-check the non-equilibrium results, we employ our new implementation of the RFDiagMC method for the computation of correlation functions in equilibrium. To do this, we first need to dertermine the diagrammatic content of the current-current correlation function that we effectively compute in our non-equilibrium calculations (in principle, in neither the $\Sigma[G_0]$ nor the $\Sigma[G]$ case will the diagrammatic content correspond to the bold perturbation theory for the current-current correlation function).
Given an approximation for the self-energy, one can express the generalized 2-particle susceptibility $\chi$ as a functional derivative of the Green's function with respect to an applied external field, $\chi = \frac{\delta G}{\delta \phi}$ \cite{Baym1961}.
In the case of the $\Sigma[G]$ approximation, this yields the self-consistent Bethe-Salpeter equation, with $\chi$ appearing on both sides of the equation. In the case of $\Sigma[G_0]$, one finds a closed expression where the non-interacting $\chi_0=G_0G_0$ appears on the r.h.s. instead.
The current-current correlation function $\Lambda$ is obtained by connecting the legs of the generalized susceptibility $\chi$ to two current vertices $v$.
We see that in the case of $\Sigma[G]$, $\Lambda$ effectively contains infinitely many skeleton diagrams of all even orders with all propagators being the full Green's functions. Up to second order, all non-zero bold skeleton diagrams are captured. However, odd orders are not captured, and at order 4 and above not all skeleton diagrams are captured.
In the $\Sigma[G_0]$ case, one obtains only three second-order diagrams, which are skeleton, but all propagators except two are bare. See Fig.\ref{fig:diagrams} and SM for details.

The $\Lambda$ diagrams from Fig.\ref{fig:diagrams} can be computed using RFDiagMC, and we denote these theories eq. $\Sigma[G]$ and eq. $\Sigma[G_0]$. The comparison with non-equilibrium results is then made by comparing $\sigma^{\mathrm{reg}}(\omega)$ and $D$. Both can be computed from $\Lambda$, namely $\mathrm{Re}\sigma^{\mathrm{reg}}(\omega\neq 0) = \frac{\mathrm{Im}\Lambda(\omega)}{\omega}$, $\sigma^{\mathrm{reg}}_\mathrm{dc} = \left.\frac{\partial \mathrm{Im}\Lambda(\omega)}{\partial \omega}\right|_{\omega\rightarrow 0}$ and $D = K - \mathrm{Re}\Lambda(\omega=0)$. The results are presented in Fig.\ref{fig:vertexcorrections}a. We see excellent agreement. In the case of $\Sigma[G]$ effective $\Lambda$ diagrams, it was enough to do only second order vertex correction diagrams to reach agreement, which means that 4th and higher order diagrams are all negligible. In the case of $\Sigma[G]$, the charge stiffness was found to be below statistical error. In the case of $\Sigma[G_0]$, the charge stiffness entirely comes from vertex corrections.

\emph{Perturbation theory for $\Lambda$.} Now that we have established the validity of our implementation, we can also use RFDiagMC to solve the perturbation theory for the current-current correlation function. We take a given self-energy approximation, construct the dressed Green's function, and then compute \emph{all} the bold skeleton diagrams, up to a given order (including the odd orders). We denote such theories as $\Lambda$-pert. with a given $\Sigma$ approximation. We find that 3rd order diagrams are practically negligible at $U=0.1$ (see SM), and the series is most likely converged already at second order. Therefore, our $\Lambda$-pert. $\Sigma[G]$ theory gives the same result as the non-eq. $\Sigma[G]$ theory. However, the $\Lambda$-pert. $\Sigma[G_0]$ approximation is different from the non-equilibrium $\Sigma[G_0]$ theory, because the vertex correction diagrams we compute are different. The results for all three distinct theories (as well as the Boltzmann theory) are compared in Fig.\ref{fig:vertexcorrections}b.

\emph{Discussion and prospects for future work.}
We observe a clear trend that $\Lambda$-pert $\Sigma[G_0]$ and $\Sigma[G]$ results become the same as $U\rightarrow 0$ (Fig.\ref{fig:vertexcorrections}b).
This indicates that the $\Lambda$-pert. series is not sensitive to the precise choice of the $\Sigma$ approximation - as our second-order $\Sigma[G]$ and $\Sigma[G_0]$ converge in the weak coupling limit, so do the corresponding low order bold-skeleton perturbation theories for $\Lambda$. However, we observe that (non-)eq. $\Sigma[G_0]$ and Boltzmann theory results remain different as $U\rightarrow 0$.

To understand this, it is important to note that the $\Lambda$ diagrams that are effectively being computed in our non-eq. $\Sigma[G_0]$ theory do not form a proper low order perturbation theory.
Even though $\Sigma[G_0]$ becomes exact as $U\rightarrow 0$ (and is even expected to perform best at low but finite coupling\cite{KozikPRL2015,GukelbergerPRB2015}, see also SM), the current response one gets from it is most likely never exact, no matter how low the value of $U$. The vertex corrections introduced this way subtract from $\sigma^\mathrm{reg}_\mathrm{dc}$, which is opposite to what is found in non-eq. $\Sigma[G]$ and the previous work with FTLM\cite{VucicevicPRL2019}. The failure of $\Sigma[G_0]$ is relevant for Ref.\cite{BergeronPRB2011} where in a similar theory, at low doping and high temperature, vertex corrections are also found to suppress dc conductivity, instead of enhance it (see SM).

On the other hand, the Boltzmann theory is equivalent to our non-eq. $\Sigma[G]$ theory, plus additional approximations. Most importantly, the Green's function appearing in the collision integral and the second-order self-energy is simplified by the quasi-particle approximation (leading to expressions formally similar to our $\Sigma[G_0]$; see SM for details). Therefore, the Boltzmann theory cannot be more accurate than our non-eq. $\Sigma[G]$ theory, and the additional approximations likely lead to the quantitatively wrong scaling we observe at $U\rightarrow 0$.

Our main finding is that the vertex corrections to dc conductivity do not vanish, even as $U\rightarrow 0$.
It appears that both the bubble and the vertex corrections diverge at small $U$ as $1/U^2$, but with a different prefactor, meaning that, as $U$ is reduced, the ratio between the bubble and the vertex corrections remains fixed. This happens despite the $U^2$ prefactor in 2nd order vert. corr. diagrams (v.c.2). The reason is that the frequency dependence $\mathrm{Im}\Lambda^{\mathrm{v.c.}2}(\omega)/U^2$ becomes singular at $\omega=0$ as $U\rightarrow 0$ (we have checked this by computing $\Lambda$ diagrams with the bare propagators, see SM).
It is possible that a similar scenario happens at higher orders as well, and that \emph{all orders} of perturbation contribute to $\sigma_\mathrm{dc}$ even at infinitesimal coupling.
Our results, however, suggest that 3rd order vert. corr. to $\sigma_\mathrm{dc}$ at $U=0.1$ are at least two orders of magnitude smaller than 2nd order.
At $U \approx 0.1-0.25$, the difference between $\Lambda$-pert. $\Sigma[G]$ and $\Lambda$-pert. $\Sigma[G_0]$ results appears to be only due to the difference in the self-energy, not due to lack of convergence of the $\Lambda$ series.

Our findings show that neither the Boltzmann theory nor the Kubo bubble are exact in the weak coupling limit. To fully confirm the strange-metal phenomenology that these two theories predict at $U\rightarrow 0$ and half-filling\cite{VucicevicPRB2023}, we need to be able to do calculations at temperatures of order $0.001-0.1$. This will require further optimizations in both our RFDiagMC and non-eq. $\Sigma[G]$ theories, which are currently limited to about $T>0.05$.
Our non-equilibrium approach can be pushed to lower temperatures by using compression methods\cite{KayeGolezSciPost2021,Kaye2023}, and the preliminary results are encouraging.
With additional optimizations outside of the scope of the current work, we should also be able to push RFDiagMC to lower temperatures and stronger coupling. The path forward is clear, at least in principle: one should attempt to converge the bare series for the equilibrium $\Sigma[G_0]$, then use it to dress the Green's function, and then try to converge the bold skeleton series for $\Lambda$.
\begin{acknowledgments}
We acknowledge useful discussions with Hugo Strand, Nenad Vukmirović, Rok Žitko, Antoine Georges, André-Marie Tremblay and Jérôme Leblanc. Computations were performed on the PARADOX supercomputing facility (Scientific Computing Laboratory, Center for the Study of Complex
Systems, Institute of Physics Belgrade). This work was granted access to the HPC resources of TGCC and IDRIS under the allocations A0170510609 attributed by GENCI (Grand Equipement National de Calcul Intensif). It has also used high performance computing resources of IDCS (Infrastructure, Données, Calcul Scientifique) under the allocation CPHT 2024. J.~V. and J.~K. acknowledge funding provided by the Institute of Physics Belgrade, through the grant by the Ministry of Science, Technological Development and Innovation of the Republic of Serbia. J.~V. and J.~K. acknowledge funding by the European Research Council, grant ERC-2022-StG: 101076100.
\end{acknowledgments}

\bibliography{refs1,refs2}
\bibliographystyle{apsrev4-1}

\newpage
\newpage
\newpage
\newpage
\begin{widetext}
\newpage


\section*{Supplementary material}

# Supplemental Material: Towards numerically exact computation of conductivity in the thermodynamic limit of interacting lattice models

Jeremija Kovačević,[1] Michel Ferrero,[2,3] and Jakša Vučičević[1]

[1]*Scientific Computing Laboratory, Center for the Study of Complex Systems, Institute of Physics Belgrade, University of Belgrade, Pregrevica 118, 11080 Belgrade, Serbia*
[2]*CPHT, CNRS, Ecole Polytechnique, Institut Polytechnique de Paris, Route de Saclay, 91128 Palaiseau, France*
[3]*Collège de France, 11 place Marcelin Berthelot, 75005 Paris, France*

(Dated: January 31, 2025)

## I. KADANOFF-BAYM EQUATIONS

Here, we briefly discuss Kadanoff-Baym (KB) equations and their numerical solution.

In order to calculate the non-equilibrium Green's function $G_\mathbf{k}$ in a given self-energy approximation $\Sigma_\mathbf{k}$ we solve the Dyson equation on the L-shaped contour $\mathcal{C}$ (Fig. 1) in the complex time plane

$$G_\mathbf{k} = G_{0\mathbf{k}} + G_{0\mathbf{k}} * \Sigma_\mathbf{k} * G_\mathbf{k} \tag{1}$$

where $*$ represents contour convolution, and $G_{0\mathbf{k}}$ is the bare Green's function given by

$$G_{0\mathbf{k}}^{-1} = \left[i\partial_t - \varepsilon_\mathbf{k}(t)\right]\delta_\mathcal{C} \tag{2}$$

where $\varepsilon_\mathbf{k}(t)$ is the time dependent dispersion relation, and $\delta_\mathcal{C}$ is the delta function on the contour $\mathcal{C}$. We assume that the eigenstates of the non-interacting Hamiltonian do not change with time. By writing the Dyson equation in terms of contour components of $G$ one obtains Kadanoff-Baym equations[1,2]

$$[-\partial_\tau - \varepsilon_\mathbf{k}(0^-)]G_\mathbf{k}^\mathrm{M}(\tau) - \int_0^\beta d\bar\tau\, \Sigma_\mathbf{k}^\mathrm{M}(\tau-\bar\tau)G_\mathbf{k}^\mathrm{M}(\bar\tau) = \delta(\tau) \tag{3}$$

$$[i\partial_t - \varepsilon_\mathbf{k}(t)]G_\mathbf{k}^\mathrm{R}(t,t') - \int_{t'}^t d\bar t\, \Sigma_\mathbf{k}^\mathrm{R}(t,\bar t)G_\mathbf{k}^\mathrm{M}(\bar t,t') = \delta(t-t') \tag{4}$$

$$[i\partial_t - \varepsilon_\mathbf{k}(t)]G_\mathbf{k}^\urcorner(t,\tau) - \int_{t_0}^t d\bar t\, \Sigma_\mathbf{k}^\mathrm{R}(t,\bar t)G_\mathbf{k}^\urcorner(\bar t,\tau) = Q_\mathbf{k}^\urcorner(t,\tau) \tag{5}$$

$$[i\partial_t - \varepsilon_\mathbf{k}(t)]G_\mathbf{k}^<(t,t') - \int_{t_0}^t d\bar t\, \Sigma_\mathbf{k}^\mathrm{R}(t,\bar t)G_\mathbf{k}^<(\bar t,t') = Q_\mathbf{k}^<(t,t') \tag{6}$$

with

$$Q_\mathbf{k}^\urcorner(t,\tau) = \int_0^\beta d\bar\tau \Sigma_\mathbf{k}^\urcorner(t,\bar\tau)G_\mathbf{k}^\mathrm{M}(\bar\tau,\tau) \tag{7}$$

$$Q_\mathbf{k}^<(t,t') = \int_{t_0}^{t'} d\bar t\, \Sigma_\mathbf{k}^<(t,\bar t)G_\mathbf{k}^\mathrm{A}(\bar t,t') - i\int_0^\beta d\tau \Sigma_\mathbf{k}^\urcorner(t,\tau)G_\mathbf{k}^\ulcorner(\tau,t'). \tag{8}$$

The complex contour $\mathcal{C}$ describes a physical setup where at the initial time $t_0$ the system is in a thermal equilibrium state. After the initial time, external fields may be applied to drive the

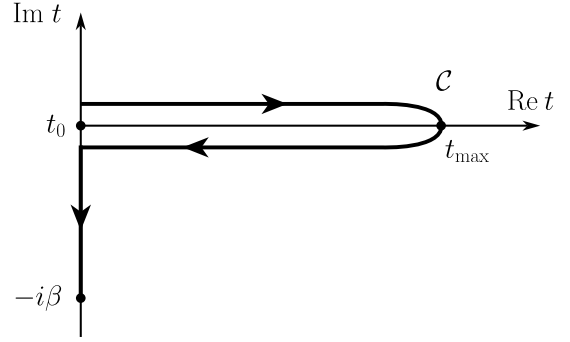

Figure 1. L-shaped contour $\mathcal{C}$ in the complex time plane.

system out of equilibrium. The evolution of the system after the initial time is described by the real parts of the contour.

Every KB equation Eq. 3 - Eq. 6 has a conjugate version, derived from the conjugate Dyson equation

$$G_\mathbf{k} = G_{0\mathbf{k}} + G_\mathbf{k} * \Sigma_\mathbf{k} * G_{0\mathbf{k}}. \tag{9}$$

Numerical solution of KB equations relies on their representation in terms of coupled Volterra integral equations of the second kind[1–3]. To do this calculation we use the NESSi package[1], and in some special cases we crosscheck the results with our own implementation. NESSi implements a version of the implicit Runge-Kuta method for solving Volterra equations where the global error scales as $\mathcal{O}(\Delta t^5)$ with size of the time step $\Delta t$. The complexity of these methods scales as $\mathcal{O}(N_t^3)$ with the number of time points $N_t$ and as $\mathcal{O}(N_\mathbf{k})$ with the number of $\mathbf{k}$-points $N_\mathbf{k}$ due to translational invariance.

## II. SELF-ENERGY

We will now discuss the calculation of the self-energy approximations used in this paper. As mentioned in the main text, we use two different self-energy approximations both represented by the expression

$$\Sigma_{ij}[G](t,t') = U^2 G_{ij}(t,t')G_{ij}(t,t')G_{ji}(t',t) \tag{10}$$

and differentiate between bold $\Sigma[G]$ and bare $\Sigma[G_0]$ approximation depending on whether the propagator used is fully dressed or bare. The fully dressed propagator is $G$ which solves the Dyson equation, while the bare propagator is $G_0$ given by Eq. 2. Since $\Sigma[G]$ is calculated using the

fully dressed propagator $G$ this makes the KB equations self-consistent and we apply an iterative procedure to solve them.

The self-consistent calculation of the bold $\Sigma[G]$ approximation can be done in two different ways. In the first approach, for one time step $(t_i, t_j)$ a loop $\Sigma_{\mathbf{k}}[G] \to G_{\mathbf{k}}[\Sigma[G]] \to G_{\mathbf{r}} \to \Sigma_{\mathbf{r}}[G] \to \Sigma_{\mathbf{k}}$ is iterated over until convergence is reached. In this case convergence is determined by the condition $\Delta G(t_i, t_j) = \|G(l) - G(l-1)\| < \epsilon$, where $l$ is the iteration index and $\epsilon$ is a small number. For different time steps $(t_i, t_j)$ and $(t'_i, t'_j)$ a different number of iterations might be needed to reach the same level of convergence. At least in principle, this allows one to reduce the overall number of computational steps.

The other possible approach, that we use in this paper, is to simply, in each iteration $l$, calculate $\Sigma_{\mathbf{k}}[G] \to G_{\mathbf{k}}[\Sigma[G]] \to G_{\mathbf{r}} \to \Sigma_{\mathbf{r}}[G] \to \Sigma_{\mathbf{k}}$ on the entire contour, until convergence is reached. In this way, the convergence criterion can be changed to reflect physically relevant quantities: in our case we are primarily interested in the current $j$, so we take $\Delta j = \|j(l) - j(l-1)\| < \epsilon$ as the condition.

For a given propagator $G_{ij}$ the calculation of the expression Eq. 10 obviously scales as $\mathcal{O}(N_t^2)$ with the number of discrete time points $N_t$, and as $\mathcal{O}(N_{\mathbf{r}}^2)$ with the number of sites $N_{\mathbf{r}}$. In order to obtain $\Sigma_{\mathbf{k}}$ from $\Sigma_{\mathbf{r}}$ and $G_{\mathbf{r}}$ from $G_{\mathbf{k}}$, we use fast Fourier transforms which scale as $\mathcal{O}(N_{\mathbf{r}} \log N_{\mathbf{r}})$. This is faster than using a $\mathbf{k}$-space expression equivalent to Eq. 10, because it scales as $\mathcal{O}(N_{\mathbf{k}}^3)$.

## III. DERIVATION OF THE BOLTZMANN EQUATION FOR LATTICE MODELS

In this section we present a derivation of the Boltzmann equation starting from Kadanoff-Baym equations for lattice models coupled to external electromagnetic fields.

Coupling (discrete) lattice models to electromagnetic fields described by gauge potentials $(\phi, \mathbf{A})$ is done by the Peierls substitution. This introduces a change in the one-particle Hamiltonian

$$H(t) = \sum_{ij,\sigma} h_{ij}(t) c_{i\sigma}^\dagger c_{j\sigma} \tag{11}$$

$$h_{ij}(t) = t_{ij} e^{i\mathbf{A}(t) \cdot (\mathbf{r}_i - \mathbf{r}_j)} + \delta_{i,j} \phi(\mathbf{r}_i) \tag{12}$$

where $t_{ij}$ is a non-local hopping amplitude and where the vector potential the vector potential $\mathbf{A}$ is assumed to be slowly varying in space.

We start from the KB equation for the lesser component of the Green's function and its conjugate version in site space

$$\sum_l [i\delta_{il} \overrightarrow{\partial}_t - h_{il}(t)] G_{lj}^<(t,t') - \sum_l \int_{t_0}^t d\bar{t}\, \Sigma_{il}^{\mathrm{R}}(t,\bar{t}) G_{lj}^<(\bar{t},t') = Q_{ij}^<(t,t') \tag{13}$$

$$\sum_l G_{il}^<(t,t') [-i\delta_{lj} \overleftarrow{\partial}_{t'} - h_{lj}(t')] - \sum_l \int_{t_0}^{t'} d\bar{t}\, G_{il}^<(t,\bar{t}) \Sigma_{lj}^{\mathrm{A}}(\bar{t},t') = \tilde{Q}_{ij}^<(t,t') \tag{14}$$

with

$$Q_{ij}^<(t,t') = \sum_l \int_{t_0}^{t'} d\bar{t}\, \Sigma_{il}^<(t,\bar{t}) G_{lj}^{\mathrm{A}}(\bar{t},t') - i \sum_l \int_0^\beta d\tau\, \Sigma_{il}^\rceil(t,\tau) G_{lj}^\lceil(\tau,t')$$

$$\tilde{Q}_{ij}^<(t,t') = \sum_l \int_{t_0}^t d\bar{t}\, G_{il}^{\mathrm{R}}(t,\bar{t}) \Sigma_{lj}^<(\bar{t},t') - i \sum_l \int_0^\beta d\tau\, G_{il}^\rceil(t,\tau) \Sigma_{lj}^\lceil(\tau,t').$$

Subtracting Eq. 14 from Eq. 13 we get

$$\begin{aligned}
i\partial_t G_{ij}^<(t,t') + i\partial_{t'} G_{ij}^<(t,t') = & \sum_l h_{il}(t) G_{lj}^<(t,t') - G_{il}^<(t,t') h_{lj}(t') \\
& + \sum_l \int_{t_0}^t d\bar{t} \Big( \Sigma_{il}^{\mathrm{R}}(t,\bar{t}) G_{lj}^<(\bar{t},t') - G_{il}^{\mathrm{R}}(t,\bar{t}) \Sigma_{lj}^<(\bar{t},t') \Big) \\
& + \sum_l \int_{t_0}^{t'} d\bar{t} \Big( \Sigma_{il}^<(t,\bar{t}) G_{lj}^{\mathrm{A}}(\bar{t},t') - G_{il}^<(t,\bar{t}) \Sigma_{lj}^{\mathrm{A}}(\bar{t},t') \Big) \\
& - i \sum_l \int_0^\beta d\tau (\Sigma_{il}^\rceil(t,\tau) G_{lj}^\lceil(\tau,t) - G_{il}^\rceil(t,\tau) \Sigma_{lj}^\lceil(\tau,t)).
\end{aligned} \tag{15}$$

We focus on the first term on the right-hand side and introduce

$$S_{ij}(t,t') = \sum_l h_{il}(t)G^<_{lj}(t,t') - G^<_{il}(t,t')h_{lj}(t') \quad (16)$$

$$= (\phi(\mathbf{r}_i) - \phi(\mathbf{r}_j))G^<_{ij}(t,t')$$
$$+ \sum_l t_{il}(t)G^<_{lj}(t,t') - G^<_{il}(t,t')t_{lj}(t') \quad (17)$$

where $t_{ij}(t) = t_{ij}e^{i\mathbf{A}(t)\cdot(\mathbf{r}_i-\mathbf{r}_j)}$.

Now we perform the Fourier transform $S_\mathbf{k} = \frac{1}{N_\mathbf{r}}\sum_{ij} S_{ij}(t,t')e^{i\mathbf{k}\cdot(\mathbf{r}_i-\mathbf{r}_j)}$, and depending on the gauge we obtain different expressions. In the $\phi$-gauge defined by $\phi(\mathbf{r}_i) = -\mathbf{E}\cdot\mathbf{r}_i$, $\mathbf{A} = 0$ we get

$$S^\phi_\mathbf{k}(t,t') = i\mathbf{E}\cdot\nabla_\mathbf{k} G^<_\mathbf{k}(t,t') \quad (18)$$

In the $A$-gauge, defined by $\phi = 0$, $\mathbf{A}(t) = -t\mathbf{E}$ we get

$$S^A_\mathbf{k}(t,t') = (\varepsilon_{\mathbf{k}+\mathbf{A}(t)} - \varepsilon_{\mathbf{k}+\mathbf{A}(t')})G^<_\mathbf{k}(t,t') \quad (19)$$

The two choices of the gauge are physically equivalent. The standard Boltzmann equation is derived in the $\phi$-gauge.

Since we are interested in finding the Boltzmann equation, i.e. the equation of motion for the non-equilibrium distribution function $\langle n_\mathbf{k}\rangle$ we concentrate on equal times $t' = t$ where

$$\left(\partial_t iG^<_\mathbf{k}(t,t') + \partial_{t'} iG^<_\mathbf{k}(t,t')\right)\Big|_{t'=t} = -\partial_t\langle n_\mathbf{k}(t)\rangle \quad (20)$$

and

$$S_\mathbf{k}(t,t) = -\mathbf{E}\cdot\nabla_\mathbf{k}\langle n_\mathbf{k}(t)\rangle. \quad (21)$$

Putting this back into the equation Eq. 15 we get

$$\partial_t n_\mathbf{k}(t) - \mathbf{E}\cdot\nabla_\mathbf{k} n_\mathbf{k}(t) = -\int_{t_0}^t d\bar{t}\Big(\Sigma^R_\mathbf{k}(t,\bar{t})G^<_\mathbf{k}(\bar{t},t) - G^R_\mathbf{k}(t,\bar{t})\Sigma^<_\mathbf{k}(\bar{t},t)$$
$$+ \Sigma^<_\mathbf{k}(t,\bar{t})G^A_\mathbf{k}(\bar{t},t) - G^<_\mathbf{k}(t,\bar{t})\Sigma^A_\mathbf{k}(\bar{t},t)\Big) \quad (22)$$
$$+ i\int_0^\beta d\tau(\Sigma^\rceil_\mathbf{k}(t,\tau)G^\lceil_\mathbf{k}(\tau,t) - G^\rceil_\mathbf{k}(t,\tau)\Sigma^\lceil_\mathbf{k}(\tau,t)).$$

where we have dropped the average brackets $\langle..\rangle$ for the sake of brevity. In order to further simplify this expression we concentrate on the limit of long times, i.e. times long after the initial state, by taking $t_0 \to -\infty$. In this limit, the mixed terms vanish. Applying this and the relations

$$G^R(t,t') = \theta(t-t')\Big(G^>(t,t') - G^<(t,t')\Big) \quad (23)$$
$$G^A(t,t') = \theta(t'-t)\Big(G^<(t,t') - G^>(t,t')\Big) \quad (24)$$

in equation Eq. 22 we get the *Boltzmann equation*

$$\partial_t n_\mathbf{k}(t) - \mathbf{E}\cdot\nabla_\mathbf{k} n_\mathbf{k}(t) = -I_\mathbf{k}(t) \quad (25)$$

where

$$I_\mathbf{k}(t) = \int_{-\infty}^t d\bar{t}\Big(\Sigma^>_\mathbf{k}(t,\bar{t})G^<_\mathbf{k}(\bar{t},t) + G^<_\mathbf{k}(t,\bar{t})\Sigma^>_\mathbf{k}(\bar{t},t)$$
$$- \Sigma^<_\mathbf{k}(t,\bar{t})G^>_\mathbf{k}(\bar{t},t) - G^>_\mathbf{k}(t,\bar{t})\Sigma^<_\mathbf{k}(\bar{t},t)\Big) \quad (26)$$

is the collision integral.

### A. Quasi-particle approximation for the collision integral in the Hubbard model

Under the assumption of a stationary state, real time components of the Green's function and the self-energy depend only on the time difference $t-t'$. The collision integral Eq. 26 becomes independent of time and can be further simplified

$$I_\mathbf{k} = \int_{-\infty}^\infty d\bar{t}\Big(G^<_\mathbf{k}(\bar{t})\Sigma^>_\mathbf{k}(-\bar{t}) - G^>_\mathbf{k}(\bar{t})\Sigma^<_\mathbf{k}(-\bar{t})\Big) \quad (27)$$

where the integration variable in equation Eq. 26 has first been shifted $\bar{t} \to t-\bar{t}$ and then for the first and third term a change of variables $\bar{t} \to -\bar{t}$ was also made.

The quasi-particle approximation now assumes that the

spectral function is a delta peak at $\omega = \varepsilon_{\mathbf{k}}$. Therefore,

$$G_{\mathbf{k}}^<(t) = \int \frac{d\omega}{2\pi} e^{-i\omega t} G_{\mathbf{k}}^<(\omega) \qquad (28)$$

$$= \int \frac{d\omega}{2\pi} e^{-i\omega t} n(\omega) A_{\mathbf{k}}(\omega) \qquad (29)$$

$$\approx \int \frac{d\omega}{2\pi} e^{-i\omega t} 2\pi \langle n_{\mathbf{k}} \rangle \delta(\omega - \varepsilon_{\mathbf{k}}) \qquad (30)$$

$$= n_{\mathbf{k}} e^{-i\varepsilon_{\mathbf{k}} t}. \qquad (31)$$

Similarly

$$G_{\mathbf{k}}^>(t) = (1 - n_{\mathbf{k}}) e^{-i\varepsilon_{\mathbf{k}} t}. \qquad (32)$$

Furthermore, if the self-energy is assumed to be given by the second order diagram $\Sigma[G]$ (as we assume in our calculations) it can be expressed in momentum space as

$$\Sigma_{\mathbf{k}}^{\lessgtr}(t) = \frac{U^2}{N^2} \sum_{\mathbf{k}', \mathbf{q}} G_{\mathbf{k}-\mathbf{q}}^{\lessgtr}(t) G_{\mathbf{k}'+\mathbf{q}}^{\lessgtr}(t) G_{\mathbf{k}'}^{\gtrless}(-t). \qquad (33)$$

Plugging this back in the collision integral Eq. 27, and applying the QP approximation we get

$$I_{\mathbf{k}}^{\text{QP},\phi} = \frac{2\pi U^2}{N^2} \sum_{\mathbf{k}', \mathbf{q}} \delta(\varepsilon_{\mathbf{k}} - \varepsilon_{\mathbf{k}-\mathbf{q}} - \varepsilon_{\mathbf{k}'+\mathbf{q}} + \varepsilon_{\mathbf{k}'})$$
$$\times \Big[ n_{\mathbf{k}}(1 - n_{\mathbf{k}-\mathbf{q}})(1 - n_{\mathbf{k}'+\mathbf{q}}) n_{\mathbf{k}'}$$
$$- (1 - n_{\mathbf{k}}) n_{\mathbf{k}-\mathbf{q}} n_{\mathbf{k}'+\mathbf{q}} (1 - n_{\mathbf{k}'}) \Big] \qquad (34)$$

which is also known as the Fermi golden rule. The equation Eq. 25 together with Eq. 34 is the starting point of the calculations in Ref. 4.

## IV. DIAGRAMMATIC CONTENT OF NON-EQUILIBRIUM THEORIES

Here we derive the diagrammatic content for the current-current correlation function that we effectively compute in our non-equilibrium theories based on a diagrammatic approximation for the self-energy. Linear response of the system to external fields $\phi$ is obtained by considering a general four-point correlation function $\chi$[5]. For convenience, we introduce a general source field $\phi_{\alpha\beta}(t, t')$ which can later be easily modified to represent a scalar or a vector potential.

All correlation functions of an interacting theory coupled to the source filed $\phi$ can be obtained from the generating functional

$$\mathcal{Z}[\phi] = \text{Tr}\Big[\mathcal{T}_\mathcal{C} e^{-i \int_\mathcal{C} dt H(t)} e^{-i \iint_\mathcal{C} dt_1 dt_2 c_{\bar{\alpha}}^\dagger(t_1) \phi_{\bar{\alpha}\bar{\beta}}(t_1, t_2) c_{\bar{\beta}}(t_2)}\Big] \qquad (35)$$

through its functional derivatives with respect to the source field $\phi$. Green's function is then defined as

$$G_{ij}^\phi(t_1, t_2) = -\frac{\delta \ln \mathcal{Z}[\phi]}{\delta \phi_{ji}(t_2, t_1)} = -i\langle \mathcal{T}_\mathcal{C} c_i(t_1) c_j^\dagger(t_2)\rangle_\phi. \qquad (36)$$

On the other hand, the Green's function can be obtained from the knowledge of the self-energy $\Sigma$, using the Dyson equation

$$G_{ij}^{-1,\phi}(t_1, t_2) = G_{0ij}^{-1}(t_1, t_2) - \phi_{ij}(t_1, t_2) - \Sigma_{ij}^\phi(t_1, t_2). \qquad (37)$$

The four-point correlation function $\chi$ is defined with the second functional derivative of $Z[\phi]$

$$\chi_{ijkl}(t_1, t_2, t_4, t_3) = -i\frac{\delta G_{ij}^\phi(t_1, t_2)}{\delta \phi_{lk}(t_4, t_3)} \qquad (38)$$
$$= -iG_{kl}^\phi(t_3, t_4) G_{ij}^\phi(t_1, t_2)$$
$$+ i\langle c_l^\dagger(t_4) c_k(t_3) c_i(t_1) c_j^\dagger(t_2)\rangle \qquad (39)$$

Using a representation of the unit operator

$$G_{i\bar{j}}^{-1,\phi}(t_1, \bar{t}_2) G_{\bar{j}k}^\phi(t_1, \bar{t}_2) = \delta_{i,k} \delta(t_1 - t_2) \qquad (40)$$

we can write

$$\chi_{ijkl}(t_1, t_2, t_4, t_3)$$
$$= -i\frac{\delta G_{i\bar{m}}^\phi(t_1, \bar{t}_5)}{\delta \phi_{kl}(t_4, t_3)} G_{\bar{m}\bar{n}}^{-1,\phi}(\bar{t}_5, \bar{t}_6) G_{\bar{n}j}^\phi(\bar{t}_6, t_2) \qquad (41)$$
$$= iG_{i\bar{m}}^\phi(t_1, \bar{t}_5) \frac{\delta G_{\bar{m}\bar{n}}^{-1,\phi}(\bar{t}_5, \bar{t}_6)}{\delta \phi_{kl}(t_4, t_3)} G_{\bar{n}j}^\phi(\bar{t}_6, t_2) \qquad (42)$$
$$= -iG_{ik}^\phi(t_1, t_4) G_{lj}^\phi(t_3, t_2)$$
$$- iG_{i\bar{m}}^\phi(t_1, \bar{t}_5) \frac{\delta \Sigma_{\bar{m}\bar{n}}^\phi(\bar{t}_5, \bar{t}_6)}{\delta \phi_{kl}(t_4, t_3)} G_{\bar{n}j}^\phi(\bar{t}_6, t_2). \qquad (43)$$

For bold $\Sigma[G]$ approximation Eq. 43 is transformed by the chain rule to

$$\chi_{ijkl}(t_1, t_2, t_4, t_3) = -iG_{ik}^\phi(t_1, t_4) G_{lj}^\phi(t_3, t_2)$$
$$+ G_{i\bar{m}}^\phi(t_1, \bar{t}_5) \frac{\delta \Sigma_{\bar{m}\bar{n}}^\phi(\bar{t}_5, \bar{t}_6)}{\delta G_{\bar{p}\bar{q}}^\phi(\bar{t}_7, \bar{t}_8)} \chi_{\bar{p}\bar{q}lk}(\bar{t}_7, \bar{t}_8, t_4, t_3) G_{\bar{n}j}^\phi(\bar{t}_6, t_2) \qquad (44)$$

which is the Bethe-Salpeter equation. The diagrammatic content of $\chi$ is then determined by the diagrammatic content of the functional derivative $\delta\Sigma/\delta G$, which is also known as the irreducible vertex in the particle-hole channel, $\Gamma^{\text{ph}}$. The diagrammatic content of $\Gamma^{\text{ph}}$ can be obtained straightforwardly if the self-energy is approximated in terms of Feynman diagrams.

In the case of a bare $\Sigma[G_0]$ approximation using the chain rule in Eq. 43 we get a different equation

$$\chi_{ijkl}(t_1,t_2,t_4,t_3) = -iG^\phi_{ik}(t_1,t_4)G^\phi_{lj}(t_3,t_2) \\ + G^\phi_{i\bar{m}}(t_1,\bar{t}_5)\frac{\delta\Sigma^\phi_{\bar{m}\bar{n}}(\bar{t}_5,\bar{t}_6)}{\delta G^\phi_{0\bar{p}\bar{q}}(\bar{t}_7,\bar{t}_8)}\chi^0_{\bar{p}\bar{q}lk}(\bar{t}_7,\bar{t}_8,t_4,t_3)G^\phi_{\bar{n}j}(\bar{t}_6,t_2) \quad (45)$$

where

$$\chi^0_{\bar{p}\bar{q}lk}(t_7,t_8,t_4,t_3) = -i\frac{\delta G^\phi_{0\bar{p}\bar{q}}(t_7,t_8)}{\delta\phi_{kl}(t_4,t_3)} = -iG^\phi_{0\bar{p}\bar{q}}(t_7,t_8)G^\phi_{0\bar{p}\bar{q}}(t_7,t_8). \quad (46)$$

Equation Eq. 45 represents a closed expression for calculating the correlation function $\chi$.

In order to obtain the optical conductivity $\sigma$ we are interested in computing the current-current correlation function defined as

$$\Lambda^{\eta,\eta'}_{\mathbf{r},\mathbf{r}'}(t,t') = -i\langle\mathcal{T}_\mathcal{C} j^\eta_\mathbf{r}(t)j^{\eta'}_{\mathbf{r}'}(t')\rangle \quad (47)$$

where $\eta$ is the spatial component of the current vector. We are especially interested in the retarded component $\Lambda^R$ which determines the linear response to the applied field through the relation

$$j^\eta_\mathbf{r}(t) = \sum_{\mathbf{r}'}\int dt' \sum_{\eta'} \Lambda^{R,\eta,\eta'}_{\mathbf{r},\mathbf{r}'}(t,t')A^{\eta'}_{\mathbf{r}'}(t') \quad (48)$$

where $A$ represents the vector potential. If the four-particle correlation function $\chi$ is known this can be done by connecting the legs of $\chi$ to the current vertex $v$. For example, if the coupling of the system to the electric field is described by an additional term

$$\sum_k \phi_k(t) v_{ijk} c^\dagger_i c_j \quad (49)$$

in the Hamiltonian, then we can calculate the dynamical response correlation function as

$$\Lambda_{ij}(t_1,t_2) = \sum_{i',i'',j',j''} v_{i,i',i''}\chi_{i',i'',j',j''}(t_1,t_1^+,t_2,t_2^+)v_{j,j',j''}. \quad (50)$$

For the explicit expression for the current vertex see Refs. 6 and 7.

## V. STUDY OF THE EFFECTS OF THE FERMIONIC BATH ON $\rho_{dc}$

In this appendix we explore the effect on $\rho_{dc}$ of coupling our model to an external fermionic bath.

We take the bath to be an infinite non-interacting system in equilibrium at temperature $T$. The full Hamiltonian of the system coupled to a fermionic bath reads

$$H = H_{\text{sys}} + H_{\text{mix}} + H_{\text{bath}} \quad (51)$$

where

$$H_{\text{mix}} = V\sum_{i,p}(c^\dagger_i b_{i,p} + c_i b^\dagger_{i,p}) \quad (52)$$

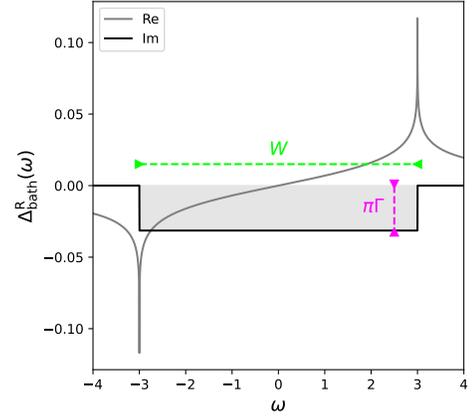

Figure 2. Our choice of the hybridization function $\Delta_{\text{bath}}(\omega)$ as defined in Eq. 57.

describes the hybridization between the system and the bath, and

$$H_{\text{bath}} = \sum_{i,p}\epsilon_p b^\dagger_{i,p} b_{i,p} \quad (53)$$

is the bath Hamiltonian. With this, the Green's function of the bath is

$$G_{\text{bath}}(t-t') = -i\sum_p \langle\mathcal{T}_\mathcal{C} b_{i,p}(t) b^\dagger_{i,p}(t')\rangle_{\text{bath}}. \quad (54)$$

As the bath is in thermal equilibrium, the Green's function is fully defined from the knowledge of the retarded Green's function, which we choose as

$$G^R_{\text{bath}}(\omega) = \frac{1}{W}\left[\ln\left|\frac{\frac{W}{2}+\omega}{\frac{W}{2}-\omega}\right| - i\pi\theta\left(\frac{W}{2}-|\omega|\right)\right]. \quad (55)$$

The bare Green's function with the additional bath is

$$\mathcal{G}^{-1}_{0\mathbf{k}}(t,t') = G^{-1}_{0\mathbf{k}}(t,t') - \Delta_{\text{bath}}(t-t') \quad (56)$$

where the hybridization function is

$$\Delta_{\text{bath}}(t-t') = V^2 G_{\text{bath}}(t-t'). \quad (57)$$

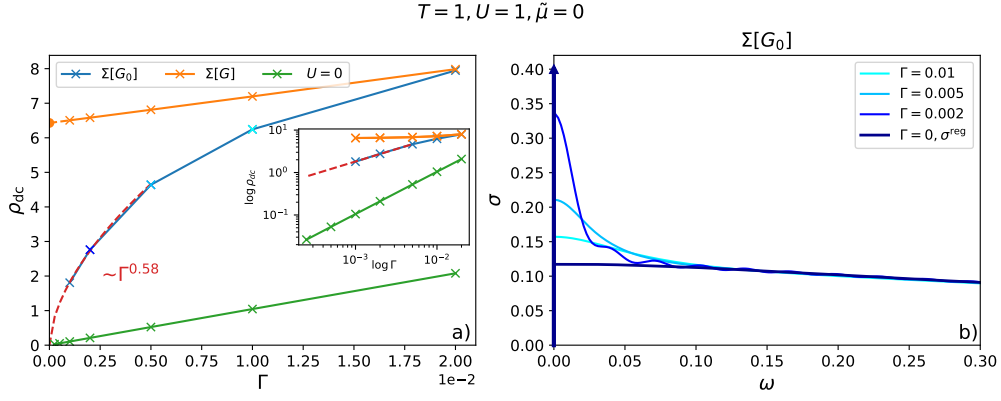

Figure 3. a) $\rho_{dc}$ vs. the bath coupling $\Gamma$ for different theories; b) $\sigma(\omega)$ for different couplings $\Gamma$ in the $\Sigma[G_0]$ approximation.

For convenience we introduce a coupling constant $\Gamma = \frac{V^2}{W}$ and keep $W$ constant. Retarded component of the hybridization function is presented in the Fig. 2.

We calculate $\rho_{dc}$ for different values of $\Gamma$ using the constant electric field protocol $A = -Et\theta(t)$. For different self-energy approximations the system behavior is drastically different, see Fig. 3.

In the case of the bold $\Sigma[G]$ approximation we can calculate the $\Gamma = 0$ point because the system reaches a stationary state, as illustrated in the main text.

On the other hand, if we use the bare $\Sigma[G_0]$ approximation, we cannot calculate $\rho_{dc}$ for $\Gamma = 0$ because the current grows indefinitely. Due to this, we compute the results at several small values of $\Gamma$ and then extrapolate to $\Gamma = 0$. The results fit to a power law form, indicating $\rho_{dc}(\Gamma = 0) = 0$, see Fig. 3a. This is in stark contrast with the $\Sigma[G]$ approximation where $\rho_{dc}$ has a finite value at $\Gamma = 0$.

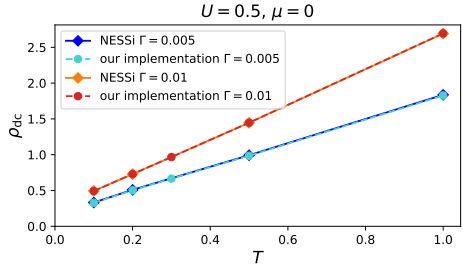

Figure 4. Comparison of $\rho_{dc}(T)$ for different implementations of Kadanoff-Baym equations.

In Fig. 3b the optical conductivity $\sigma(\omega)$ in the bare approximation is presented. We explore the $\Gamma \to 0$ limit and we see that $\sigma$ develops a delta peak at $\omega = 0$ for $\Gamma = 0$, elucidating the presence of charge stiffness in this approximation.

As an additional check, we created our own implementation of the Kadanoff-Baym equations for this setup and compared it to the NESSi implementation (Fig. 4) and the results are in excellent agreement. In our implementation the initial thermal state is at $T = \infty$. This approach is possible because the system gets cooled down to temperature $T$ due to the coupling with the bath, and indeed we observe a stationary state in our results.

In practice, because the initial state is at infinite temperature, $\beta = 0$, the imaginary part of the contour drops out, as well as all the mixed terms. This leaves us with only two KB equations, one for $G^<$ and the other one for $G^>$. The system then evolves to a stationary state at temperature $T$ due to the coupling with the bath. We solved KB equations using the implicit Runge-Kutta method of the 4-th order where the global error scales as $\mathcal{O}(\Delta t^4)$. NESSi has better error scaling $\mathcal{O}(\Delta t^5)$, this did not make a difference in the final results.

## VI. INVERSE LINEAR RESPONSE THEORY

In this appendix we will discuss our approach to using non-equilibrium protocols for calculating $\rho_{dc}$.

As mentioned in the main text, we use three different non-equilibrium protocols for calculating $\rho_{dc}$. They are presented in Table I.

Table I. Non-equilibrium protocols to extract $\sigma(\omega)$

| | | | |
|---|---|---|---|
| a) | Constant electric field | $A = -Et$ | $\sigma_{dc} = \langle j(t \to \infty)\rangle/E$ |
| b) | Short pulse of elec. field | $A = -a\theta(t)$ | $\sigma(t) = \langle j(t)\rangle/a$ |
| c) | Short pulse of vec. pot. | $A = -a\delta(t)$ | $\Lambda^R(t) = \langle j(t)\rangle/a$ |

We consider two equivalent linear response equations for the uniform current[8,9]

$$\langle j(t)\rangle = \int_{-\infty}^{t} dt'\, \sigma(t-t') E(t') \quad (58)$$

$$\langle j(t)\rangle = \int_{-\infty}^{t} dt'\, \Big(K\delta(t-t') - \Lambda(t-t')\Big) A(t') \quad (59)$$

with $K = \frac{\langle E_{\text{kin}}\rangle}{2}$ being the average kinetic energy per particle per spatial dimension and $E$ being the electric field. If we apply protocol b) to Eq. 58 and protocol c) to Eq. 59 we get

$$\langle j(t)\rangle = a\sigma(t) \quad (60)$$
$$\langle j(t)\rangle = a\Lambda(t). \quad (61)$$


After solving KB Eqs. 3-6 in a given protocol we get the current $j(t)$, which we then use in Eqs. 60 and 61 to obtain the correlation functions. In this way, we have *inverted* the linear response Eqs. 58 and 59.

In the case of the constant electric field protocol a) we assume a stationary state, and we can write

$$\langle j(t \to \infty) \rangle = E \int_{-\infty}^{\infty} dt' \sigma(t - t') \quad (62)$$
$$= E \sigma_{\text{dc}}. \quad (63)$$

## VII. BENCHMARKS

Here we will discuss the dependence of the results on the numerical parameters.

Fig. 5 shows the dependence of the calculated current $\langle j(t) \rangle$ on the linear lattice size $L$ and temperature $T$ in different approximations in the b) protocol. We see the current displays large finite size effects for both approximations.

does appear noticeably different; the reason is that there is a cancellation of difference between shorter and longer times.

In the $\Sigma[G_0]$ approximation we also see "revivals" occuring for a long time after the field pulse, while in the $\Sigma[G]$ approximation no revivals are observed at any lattice size $L$.

Fig. 6a shows the current $\langle j(t) \rangle$ in the bare $\Sigma[G_0]$ approximation calculated using the b) protocol for different time steps $\Delta t$.

Fig. 6b and 6c present $\langle j(t) \rangle$ in $\Sigma[G]$ approximation calculated using the b) and a) protocols for two different values of the parameters $a$ and $E$, respectively. Fig. 6d shows $\langle j(t) \rangle$ in the c) protocol for different durations of the applied pulse $\Delta t_{\text{pulse}}$. Note that in the b protocol we do not have a finite duration of the applied field, but rather exactly $\Delta t_{\text{pulse}} = 0$; this means we have no additional numerical parameter to control the results with respect to.

In conclusion, the calculated currents do not depend significantly on the numerical parameters and we have reached a linear regime.

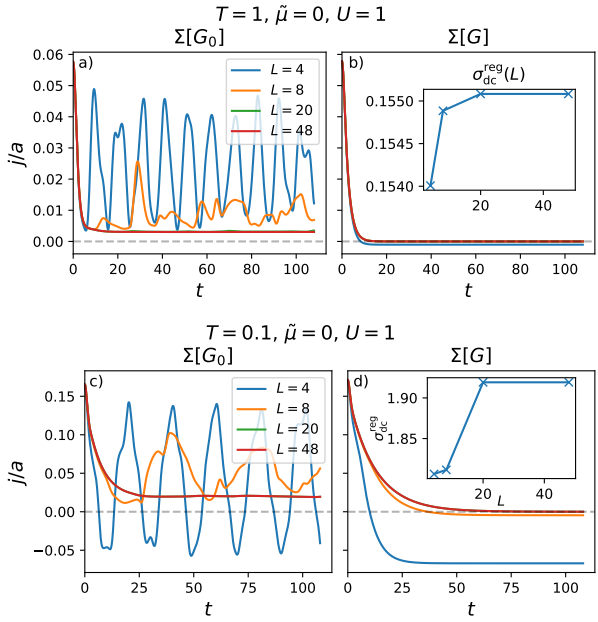

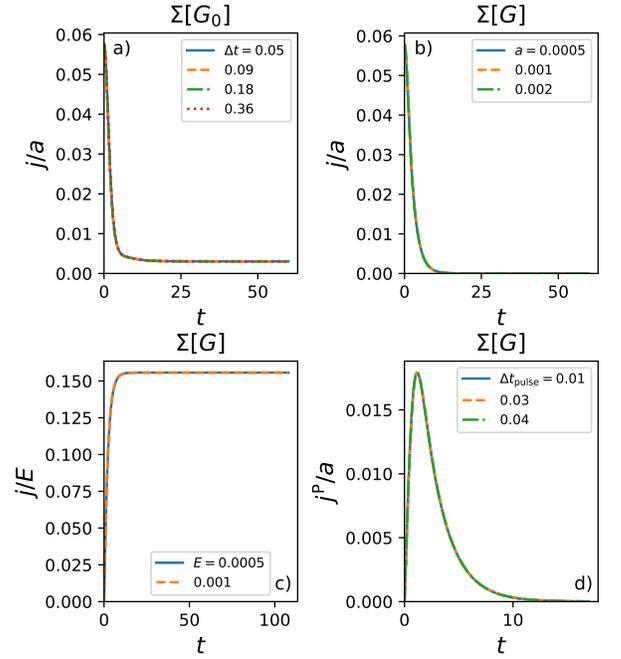

Figure 5. $j/a$ in the b) protocol as a function of the linear lattice size $L$ in different self-energy approximations and at different temperatures.

In the $\Sigma[G_0]$ approximation, at a higher temperature $T = 1$ in Fig. 5a we can see that the current is strictly positive for all lattice sizes, while this is not necessarily the case for the lower temperature $T = 0.1$ (Fig. 5c). Also, it can be seen that the charge stiffness does not decrease with increasing lattice size.

In the $\Sigma[G]$ approximation the current changes sign due to finite size effects regardless of the temperature, see Figs. 5b and 5d. We see that in this case charge stiffness decreases with linear lattice size $L$. In the insets of Fig.5 we show $\sigma_{\text{dc}}^{\text{reg}}$ which is obtained as $\int dt (j(t) - D)$. One can see that finite size effects are not strong in terms of $\sigma_{\text{dc}}^{\text{reg}}$, even though $\sigma(t) - D$

Figure 6. a) $j/a$ in the b) protocol for different time steps $\Delta t$; b) $j/a$ in the b) protocol for different values of the parameter $a$; c) $j/E$ in the a) protocol for different $E$; d) $j/a$ in the c) protocol for different durations of the applied pulse $\Delta t_{\text{pulse}}$.

## VIII. LOCAL SELF-ENERGY

In the case of local approximations of the Luttinger-Ward functional, the irreducible vertex $\delta\Sigma/\delta G$ featured in the Bethe-Salpeter equation Eq. 44 is fully local. On lattices that preserve inversion symmetry this ensures that the vertex corrections for the uniform current-current correlation function $\Lambda$ cancel out[6,10]. Therefore, as an additional check, we compare

non-equilibrium and Kubo bubble values of $\rho_{dc}$ in the local self-energy $\Sigma[G_0^{\rm loc}]$ approximation in Fig. 7. We see excellent agreement between these theories.

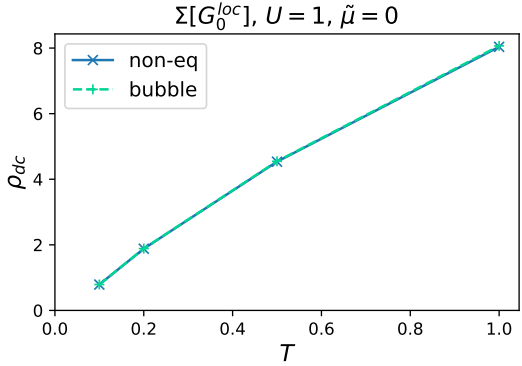

Figure 7. Comparison of $\rho_{\rm dc}(T)$ for the non-equilibrium $\Sigma[G_0^{\rm loc}]$ approximation and the bubble approximation.

## IX. CHARGE STIFFNESS IN $\Sigma[G_0]$ APPROXIMATION

In this section we discuss the charge stiffness $D$ that is present in the non-equilibrium $\Sigma[G_0]$ approximation. When

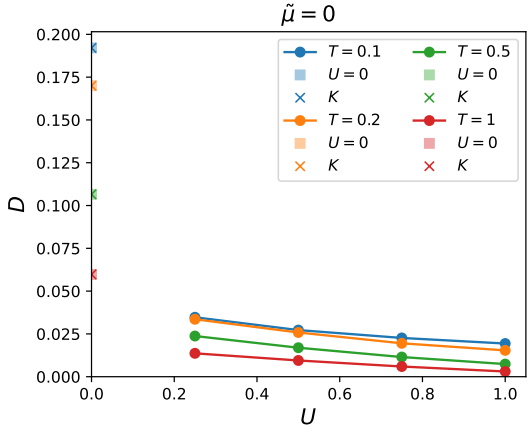

Figure 8. Charge stiffness $D$ in the $\Sigma[G_0]$ approximation. The $U = 0$ point is calculated using the short pulse of electric field protocol. $K$ is calculated according to Eq. 67. Colors on the plot indicate the temperature $T$.

the charge stiffness develops we can write the conductivity $\sigma$ as

$$\sigma(t) = D + \sigma^{\rm reg}(t) \quad (64)$$

where $\sigma^{\rm reg} \to 0$ as $t \to \infty$. On the other hand, the current-current correlation function $\Lambda$ sees charge stiffness only through the optical sum rule

$$\int_0^\infty dt\, \Lambda(t) = K - D. \quad (65)$$

This is due to the fact that $\sigma$ and $\Lambda$ are related by the derivative of $\sigma$

$$\partial_t \sigma(t) = \partial_t \sigma^{\rm reg}(t) = -\Lambda(t) \quad (66)$$

which sees only the regular part of conductivity $\sigma^{\rm reg}$.

Charge stiffness develops at all the temperatures and at all the couplings we examined, Fig. 8. We see that the charge stiffness increases with smaller couplings, and at $U = 0$ we have $D = K$. This is clearly understood, by taking into account that at $U = 0$, $\Lambda(\omega) = 0$ (see Section XIII), and therefore the entire optical conductivity $\sigma(\omega)$ is contained in the charge stiffness peak at $\omega = 0$. We confirm that, at $U = 0$, $D = K$, by looking at results of protocol b) with $\Sigma = 0$, and comparing to $K$ computed as

$$K = \frac{\langle E_{\rm kin}(U=0) \rangle}{2} = \sum_{\mathbf{k}} n_F(\varepsilon_{\mathbf{k}}) \varepsilon_{\mathbf{k}}. \quad (67)$$

In the b) protocol, charge stiffness manifests as a non-physical stationary state with constant current. The density distribution in the Brillouin zone $\langle n_{\mathbf{k}} \rangle$ in this state is different from the initial equilibrium one, as can be seen in Fig. 9. One can see that it does not have the full symmetry of the lattice. Due to this, the contribution to the uniform current of different $\mathbf{k}$-points does not cancel out.

## X. $\Sigma[G](\omega)$ AND $\Sigma[G_0](\omega)$ COMPARISON

Here we compare two different self-energy approximations used in the paper. We focus on thermal equilibrium state and compute $\Sigma[G_0]$ using a separate well-optimized code implemented in frequency domain[11]. We compute $\Sigma[G]$ by solving the self-consistent KB equations, but without any external fields. The resulting $\Sigma$ is then Fourier transformed from time domain to frequency domain. In Fig. 10 a comparison of the bare $\Sigma[G_0]$ and the bold $\Sigma[G]$ approximations is presented for two different temperatures. In Fig. 11 we compare two different approximations for different values of $U$. In the limit $U \to 0$ we see that $\Sigma[G] \to \Sigma[G_0]$. We also see that $\Sigma[G]$ has in general a smoother frequency dependence and longer high-frequency tails.

## XI. LIMITATIONS OF SELF-ENERGY APPROXIMATIONS

In this section we will discuss some limitations of self-energy approximations that we used.

We expect both $\Sigma[G]$ and $\Sigma[G_0]$ approximations to be valid only for weak couplings. In Ref. 12 and Ref. 13 authors show how for moderate couplings, our second order $\Sigma[G_0]$ is found to be in better agreement with numerically exact results than the second order $\Sigma[G]$. However, we document here that there is a hard cutoff value of $U$ beyond which $\Sigma[G_0]$ starts giving unphysical results for the Green's function. By contrast, the $\Sigma[G]$ approximation does not have such a limitation, albeit the

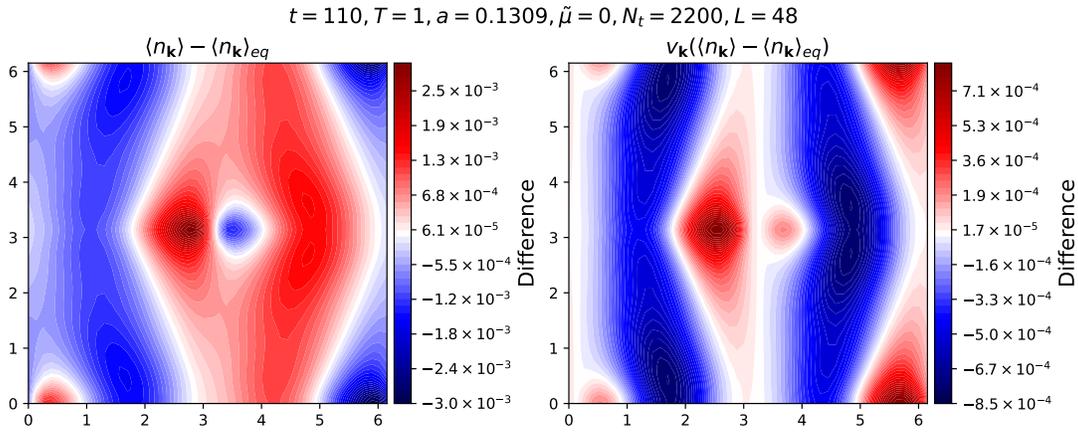

Figure 9. Characterization of the unphysical stationary state in case of the bare $\Sigma[G_0]$ approximation.

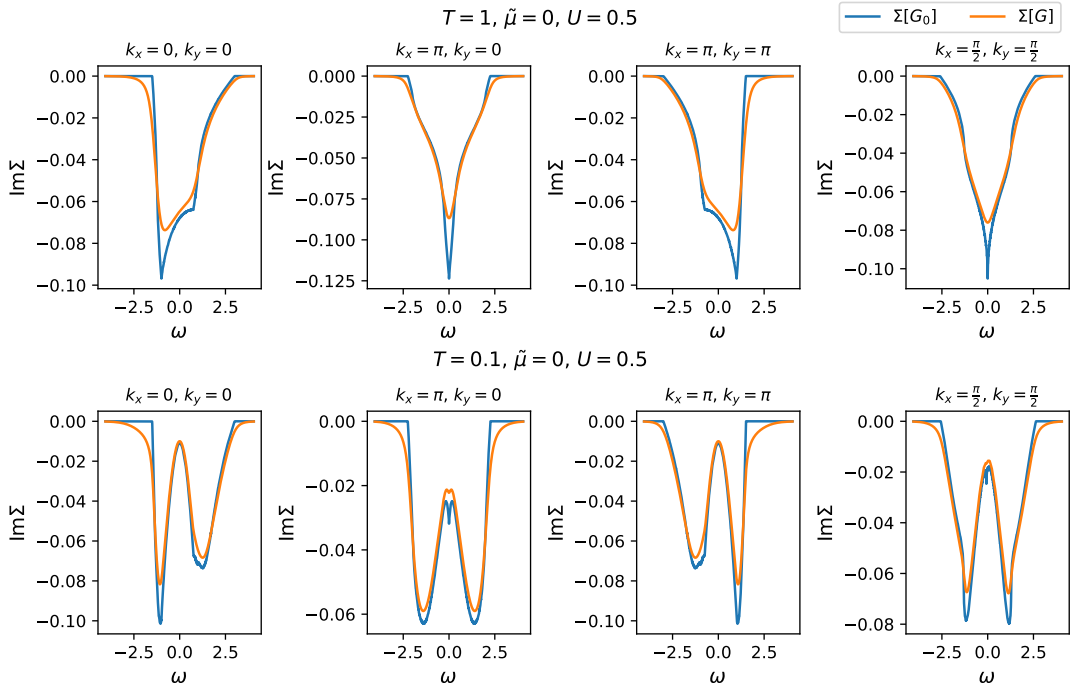

Figure 10. Comparison of different self-energy approximations for two different temperatures.

convergence of the iterative $\Sigma[G]$ solution is not guaranteed at strong couplings.

The limitation of the $\Sigma[G_0]$ approximation comes from the fact that $\text{Im}\Sigma(\omega)$ always goes to exactly zero at some finite (and not very large) frequency. At large $U > U_{\max}$ this leads to the appearance of unphysical coherent states at certain wave vectors. Namely, if we look at the Green's function

$$G(\omega) = \frac{1}{\omega + \mu - \varepsilon_{\mathbf{k}} - U^2 \text{Re}\Sigma_{\mathbf{k}}(\omega) - iU^2 \text{Im}\Sigma_{\mathbf{k}}(\omega)} \quad (68)$$

the real part of the denominator can have zeros $\omega^*$ such that $\text{Im}\Sigma(\omega^*) = 0$, yielding a delta peak structure in the Green's function, which would correspond to an infinitely long lived state. This can lead to infinite conductivity, even if there is no charge stiffness. Fig. 12b illustrates on one example the real part of the denominator of $G$ and its root $\omega^*$. In can be seen that the root $\omega^*$ is at a frequency where $\text{Im}\Sigma(\omega^*) = 0$. This is in contrast with the smaller value of $U$, see Fig. 12a, where there are no roots $\omega^*$ for which $\text{Im}\Sigma(\omega^*)$ vanishes. In Fig. 12c $U_{\max}$ is presented as a function of temperature for different chemical potentials $\tilde{\mu}$. We conclude that, regardless of the accuracy of the second order $\Sigma[G_0]$ approximation in terms of describing the correct $\Sigma$ at low frequency, it cannot be safely used for the computation of spectral and transport properties beyond about $U \approx 1.3$.

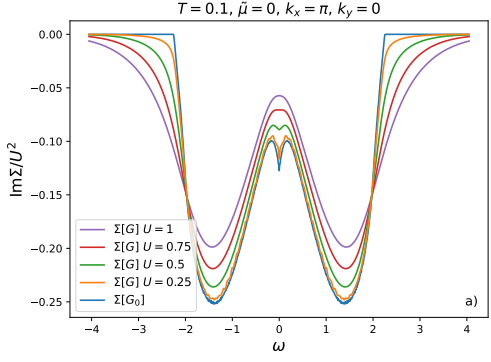

Figure 11. Comparison between $\Sigma[G_0]$ and $\Sigma[G]$ for different couplings $U$.

## XII. REAL-FREQUENCY DIAGRAMMATIC MONTE CARLO (RFDIAGMC)

The basis for our RFDiagMC calculations is explained in Ref. 14. The idea is that imaginary-time integrals in a general Feynman diagram $D$ can be solved analytically, to arrive at a general expression of the form:

$$D(i\omega_n) = \sum_{j,p} \frac{\mathcal{A}_{jp}}{(i\omega_n - \mathcal{E}_j)^p} \tag{69}$$

The analytical continuation is then trivially performed, by replacing the Matsubara frequency by the complex frequency $i\omega_n \to z$. We can then take $z = \omega + i0^+$ to retrieve the contribution of the diagram to the retarded correlation function, $D^{\mathrm{R}}(\omega)$. By binning the pole amplitudes on a uniform real-frequency grid, we can ultimately avoid using any broadening. The only numerical parameter in the calculation is the frequency resolution. In practice we do the calculation with a very dense grid, and a posteriori we do a resampling procedure, where we merge adjacent bins until we obtain smooth results.

We have performed two important generalizations of the algorithm, compared to our work in Ref. 14. First, our algorithm is now general for an arbitrary physical quantity, which allows us to compute both the self-energy diagrams and the current-current correlation function diagrams. Second, the propagators in the diagram can now be the dressed Green's functions, as is necessary for computation of bold skeleton diagrams. Our algorithm can work with fully general propagators of the form:

$$G(\tau) = \eta s_\tau \int \mathrm{d}\omega \rho(\omega) n_\eta(-s_\tau \omega) e^{-\omega \tau} \tag{70}$$

where $\eta = \pm 1$ for bosons/fermions, $s_\tau$ is the sign of $\tau \in [-\beta, \beta]$, with $\beta$ the inverse temperature, and $\rho(\omega) = -\frac{1}{\pi}\mathrm{Im}G^{\mathrm{R}}(\omega)$ is the corresponding spectral function, obtained from the retarded Green's function $G^{\mathrm{R}}$. In practice, this means that in the expressions that are being evaluated, now we

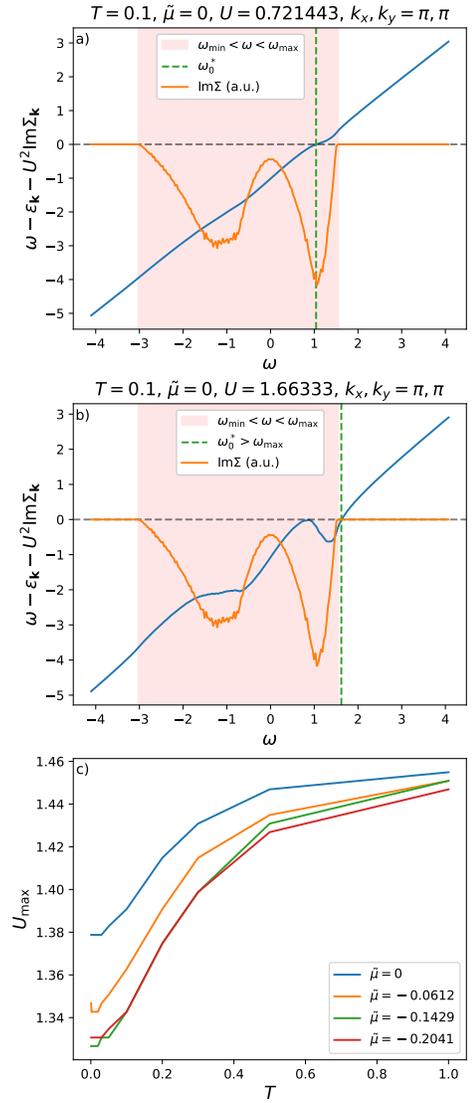

Figure 12. a) Real part of the denominator of $G$ and imaginary part of the self-energy for $U < U_{\max}$; b) Real part of the denominator of $G$ and imaginary part of the self-energy for $U > U_{\max}$; c) $U_{\max}$ at different temperatures and chemical potentials. On all plots $\mathrm{Im}\Sigma$ is given in arbitrary units (a.u.).

have additional frequency integrals to be solved numerically. At this point, one must consider whether it is better to solve the diagrams directly on the three-piece contour. When solving bold skeleton diagrams using our approach, one essentially replaces a smaller number of time-integrals by a larger number of frequency integrals. However, the frequency integrals appear easier to solve. The real-time integrands are highly oscillatory, and the integration domain is infinite. Real-time integrations are a difficult problem that has not yet been solved in the lattice context. By contrast, the frequency integrals do not contribute to the sign problem as $\rho(\omega)$ is positive definite (as long as we work in momentum basis) and the integration domain is very restricted ($\rho(\omega)$ is sharply peaked if the self-energy used to dress the propagator is small; the prob-

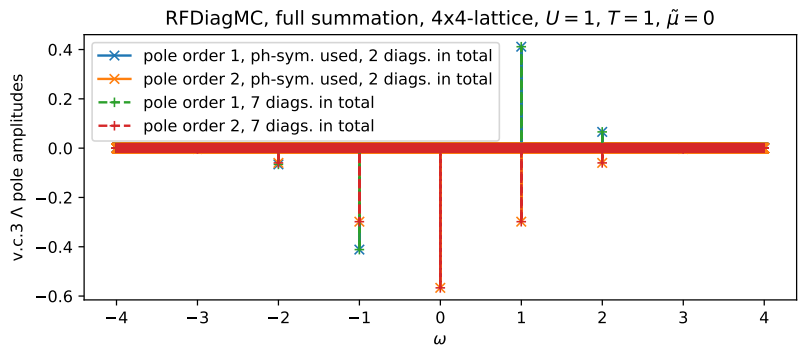

Figure 13. Benchmark of our algorithm for identifying the minimal set of diagrams: The results shown are the third order vertex corrections, computed for a small lattice, using the bare propagator and the full momentum summation to eliminate any statistical errors. The results are computed with and without making use of the particle-hole symmetry which reduces the number of diagrams from 7 to 2 and show perfect agreement.

lem thus becomes easier at weaker coupling). By using the product of spectral functions $\rho(\omega)$ of all the propagators in a diagram as Monte Carlo weight, we can successfully perform the frequency integration.

In practice, we use Sobol sequence quasi Monte Carlo for the summation over inner momenta $\mathbf{k}$, and for each given choice of momenta, we perform a weighted Monte Carlo over frequencies.

It turns out that it is very important that the momentum integrations are performed in a continuous BZ, i.e. in the true thermodynamic limit. We have shown in our previous work[14] that finite-size lattice calculations lead to noisy structures in frequency spectra. We have observed similar behavior also in the context of our current-current correlation function calculations - yet, here we are here interested in the derivative of $\text{Im}\Lambda(\omega)$ at $\omega = 0$, where this function is nearly singular, and rapidly changes slope in a very narrow frequency window. Any noise coming from finite lattice size could easily affect our $\sigma_{\text{dc}}$ estimates.

In practice, we compute the self-energy on a large lattice, and then interpolate it linearly on-the-fly in the $\Lambda$ calculation, as needed to construct the dressed propagator $G$ that enters the diagram. In the case of the $\Sigma[G]$ approximation, we have computed the self-consistent self-energy in real time (in absence of any field, with the lattice size $48 \times 48$), and then performed the Fourier transform to real frequency. In the case of the $\Sigma[G_0]$, we have performed the calculation in frequency using a separate, highly accurate and efficient code we had developed previously[11] (with the lattice size $64 \times 64$).

## XIII. VERTEX CORRECTIONS

In this section we discuss and show raw results for the bubble and vertex correction contributions to the current-current correlation function $\Lambda(\omega)$. The calculation yields the imaginary part, and we reconstruct the real-part using Kramers-Kronig relations (in particular we need $\text{Re}\Lambda(\omega = 0)$ for the estimate of charge stiffness).

The calculation of vertex correction diagrams is expensive. We have used symmetries to reduce the overall number of diagrams to be computed. In general, the current-vertex symmetry $v_{\mathbf{k}} = -v_{-\mathbf{k}}$ will make contributions of certain diagrams zero. Due to this symmetry all first order vertex correction diagrams drop out. The bosonic symmetry of the correlation function itself $\Lambda(\tau) = \Lambda(-\tau)$ means that diagrams which are topologically the same up to the exchange of terminals (current vertices) will yield the same contribution, and it is sufficient to compute only one of them. Finally, at half-filling, one can use particle-hole symmetry to further reduce the number of diagrams[15]. However, this will only hold for fully dressed diagrams (not possible in the case of partially dressed diagrams as one effectively obtains in our non-eq. $\Sigma[G_0]$ theory). At half-filling, in total, we need to compute a single diagram at order 2 (out of 4 possible diagrams), and 2 diagrams at order 3 (out of 11 possible diagrams).

To check the validity of our simplifications, we have performed calculations of all diagrams and compared to the results obtained using the minimal set of diagrams, on examples which are not numerically expensive. In Fig. 13 we show the calculation of 3rd order vertex correction diagrams with the bare-propagators, on a small $4 \times 4$ lattice, with and without the use of ph-symmetry, and show how the results agree perfectly.

In Fig. 14 we show our main $\text{Im}\Lambda(\omega)$ results. It is clear that as $U$ is reduced, the bubble vanishes, but at the same time the slope at $\omega = 0$ diverges. Ultimately, if we take exactly $U = 0$ and compute the bubble with the bare propagators, we get $\Lambda(\omega) = 0$, as already mentioned in Appendix IX.

Similarly, the vertex correction diagrams become more and more singular around $\omega = 0$ as $U \to 0$. We have checked that, when vertex correction diagrams are computed with the bare propagator, one observes a true discontinuity (jump) in $\text{Im}\Lambda(\omega)$ at $\omega = 0$, see Fig. 15, which means that contributions of individual diagrams to $\sigma_{\text{dc}}$ become of the form $0 \times \infty$ as $U \to 0$. If for diagrams at order $k$, the divergence is $1/U^{k+2}$ or faster, these vertex correction diagrams will contribute even at infinitesimal coupling. Of course, there is an overall prefactor which is expected to decrease with increasing $k$, and the contribution of high orders may be small and even negligible.

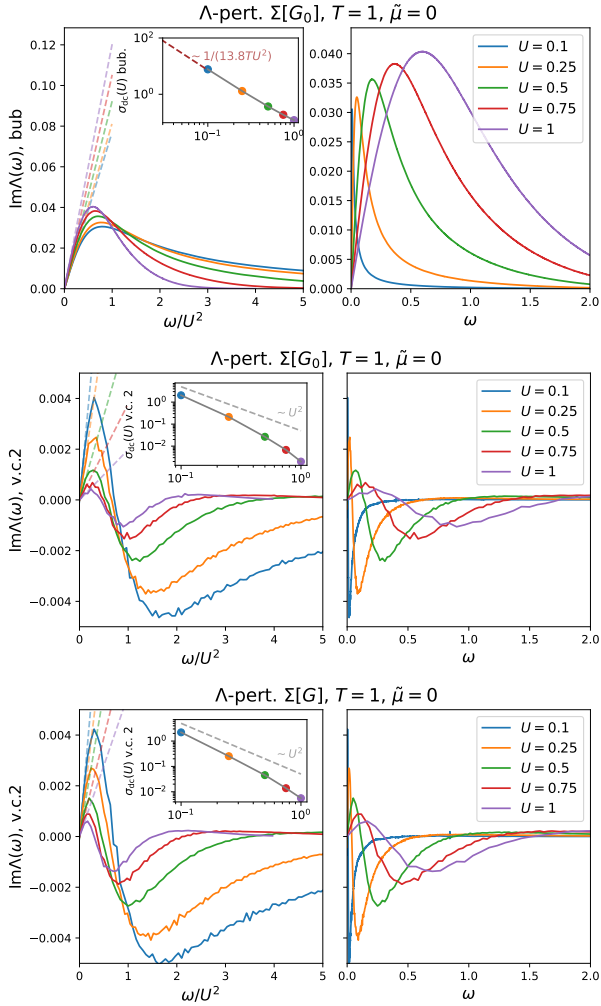

Figure 14. Raw current-current correlation function results for different values of $U$. First row: Kubo bubble contribution, second and third rows: second-order vertex corrections computed with the propagator dressed with two different self-energy approximations. In the left column, frequency axis has been rescaled by $U^2$ for easier inspection. The $\sigma_{\rm dc}$ contribution is the angle at $\omega = 0$, indicated by the dashed line, and shown as a function of $U$ in the insets.

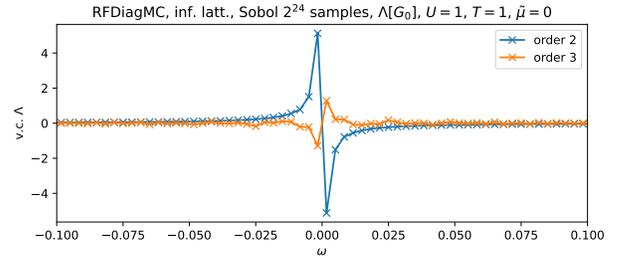

Figure 15. Vertex corrections to the current-current correlation function at second and third order, computed using bare propagators. There is a discontinuity (jump) at $\omega = 0$, which yields an infinite contribution to $\sigma_{\rm dc}$. This indicates that the bare series for $\Lambda(\omega \approx 0)$ is not convergent.

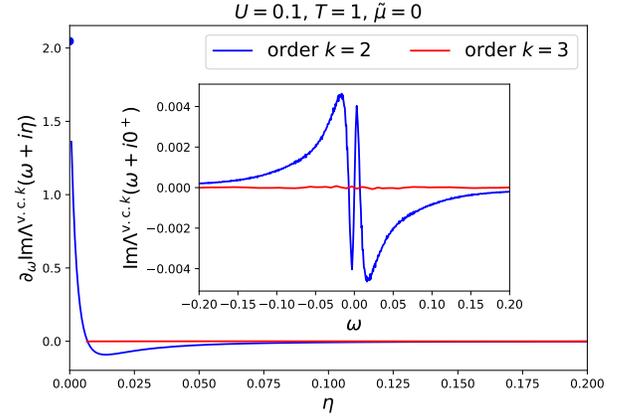

Figure 16. Comparison of the second and third order vertex corrections, computed using propagators dressed with $\Sigma[G_0]$. Inset: frequency dependence of the imaginary part. Main panel: contribution to $\sigma_{\rm dc}$ as a function of the broadening $\eta$. Third order contribution to $\sigma_{\rm dc}$ appears to be below the statistical error bar.

The calculation of 3rd order vertex corrections to high enough precision to extract $\sigma_{\rm dc}$ is currently not possible. We find that $\mathrm{Im}\Lambda^{\rm v.c.3}(\omega)$ is overall 2 orders of magnitude smaller than $\mathrm{Im}\Lambda^{\rm v.c.2}(\omega)$. We believe it is highly unlikely that v.c. of 3rd order (and higher) contribute significantly. Further optimizations in our code will be necessary to fully resolve this question.


\end{widetext}
\end{document}